\documentclass[fleqn,usenatbib]{mnras}

\usepackage{newtxtext,newtxmath}

\usepackage[T1]{fontenc}

\DeclareRobustCommand{\VAN}[3]{#2}
\let\VANthebibliography\thebibliography
\def\thebibliography{\DeclareRobustCommand{\VAN}[3]{##3}\VANthebibliography}

\usepackage{graphicx}	
\usepackage{amsmath}	
\usepackage{multirow} 

\newcommand{\Msun}{{~\rm M_\odot}}
\newcommand{\ergs}{{~\rm erg\cdot s^{-1}}}
\newcommand{\Msunh}{{~\rm M_\odot/h}}
\newcommand{\mpch}{~\rm Mpc/h}
\newcommand{\lgalaxies}{\textsc{L-Galaxies}}

\title[Simulating ELGs]{Simulating emission line galaxies for the next generation of large-scale structure surveys}

\author[Pei et al.]{
Wenxiang Pei,$^{1,2,3}$
Qi Guo,$^{1,2,3}$\thanks{E-mail: guoqi@nao.cas.cn}
Ming Li,$^{1}$
Qiao Wang,$^{1}$
Jiaxin Han,$^{4,5,6}$
Jia Hu,$^{1,3}$
Tong Su,$^{1,2,3}$
\newauthor
Liang Gao,$^{1,7,2,3}$
Jie Wang,$^{1,2,3}$
Yu Luo$^{8,9}$
and Chengliang Wei$^{8}$
\\
$^{1}$Key Laboratory for Computational Astrophysics, National Astronomical Observatories, Chinese Academy of Sciences, Beijing 100101, China\\
$^{2}$Institute for Frontiers in Astronomy and Astrophysics, Beijing Normal University, Beijing 102206, China\\
$^{3}$School of Astronomy and Space Science, University of Chinese Academy of Sciences, Beijing 10049, China \\
$^{4}$Department of Astronomy, Shanghai Jiao Tong University, Shanghai 200240, China\\
$^5$Key Laboratory for Particle Astrophysics and Cosmology (MOE), Shanghai 200240, China\\
$^6$Shanghai Key Laboratory for Particle Physics and Cosmology, Shanghai 200240, China\\
$^{7}$School of Physics and Microelectronics, Zhengzhou University, Zhengzhou, 450001, China\\
$^{8}$Purple Mountain Observatory, Chinese Academy of Sciences, Nanjing 210008, China\\
$^{9}$National Basic Science Data Center, Zhongguancun South 4th Street, Beijing 100190, China\\
}

\date{Accepted XXX. Received YYY; in original form ZZZ}

\pubyear{2015}

\begin{document}
\label{firstpage}
\pagerange{\pageref{firstpage}--\pageref{lastpage}}
\maketitle

\begin{abstract}

We investigate emission line galaxies across cosmic time by combining the modified L-Galaxies semi-analytical galaxy formation model with the JiuTian cosmological simulation. We improve the tidal disruption model of satellite galaxies in L-Galaxies to address the time dependence problem. We utilise the public code CLOUDY to compute emission line ratios for a grid of HII region models. The emission line models assume the same initial mass function as that used to generate the spectral energy distribution of semi-analytical galaxies, ensuring a coherent treatment for modelling the full galaxy spectrum. By incorporating these emission line ratios with galaxy properties, we reproduce observed luminosity functions for H$\alpha$, H$\beta$, [OII], and [OIII] in the local Universe and at high redshifts. We also find good agreement between model predictions and observations for auto-correlation and cross-correlation functions of [OII]-selected galaxies, as well as their luminosity dependence. The bias of emission line galaxies depends on both luminosity and redshift. At lower redshifts, it remains constant with increasing luminosity up to around $\sim 10^{42.5}\ergs$ and then rises steeply for higher luminosities. The transition luminosity increases with redshift and becomes insignificant above $z$=1.5. Generally, galaxy bias shows an increasing trend with redshift. However, for luminous galaxies, the bias is higher at low redshifts, as the strong luminosity dependence observed at low redshifts diminishes at higher redshifts. We provide a fitting formula for the bias of emission line galaxies as a function of luminosity and redshift, which can be utilised for large-scale structure studies with future galaxy surveys.

\end{abstract}




\section{Introduction}

In the past few decades, large-scale surveys have propelled revolutionary developments in astronomy. Various surveys, such as the Sloan Digital Sky Survey \citep[SDSS,][]{2000AJ....120.1579Y, 2006AJ....131.2332G}, the Dark Energy Survey \citep[DES,][]{2016MNRAS.460.1270D}, and the Hyper Suprime-Cam Subaru Strategic Program \citep[HSC-SSP,][]{2018PASJ...70S...8A,2018PASJ...70...66K} have significantly enhanced our understanding of the universe and the underlying theories of galaxy formation. These surveys have contributed to the solidification of the $\Lambda$CDM cosmological framework, precise measurements of the expansion rate of the universe, and the provision of extensive data on the large-scale structure of the universe, including the mapping of millions of galaxies. In addressing evolving scientific challenges, currently ongoing and upcoming large-scale next-generation surveys, such as the Chinese Space Station Telescope \citep[CSST,][]{2011SSPMA..41.1441Z}, the Dark Energy Spectroscopic Instrument \citep[DESI,][]{2022AJ....164..207D}, \textit{Euclid} \citep{2011arXiv1110.3193L}, the Large Synoptic Survey Telescope \citep[LSST,][]{2019ApJ...873..111I}, the \textit{Nancy Grace Roman Space Telescope} \citep[\textit{Roman,}][]{2012arXiv1210.7809D,2013arXiv1305.5425S}, and the Subaru Prime Focus Spectrograph \citep[PFS,][]{2014PASJ...66R...1T,2016SPIE.9908E..1MT} aim to achieve substantial breakthroughs in fundamental issues such as the origin and evolution of dark matter and dark energy, as well as the origin and evolution of galaxies and black holes. These state-of-art instruments and surveys have lower detection limits and better spatial resolution, thereby extending observational data towards fainter objects and higher redshifts. 

Emission line galaxies (ELGs) serve as the main tracer at z $\sim$ 1 for large-scale surveys, playing a crucial role in estimating photometric and spectroscopic redshifts and properties related to galaxy formation. In recent years, numerous theoretical works have also focused on the formation of ELGs. Some studies have utilised the Halo Occupation Distribution (HOD) method to assign emission line luminosities to dark matter-only simulations \citep{2012MNRAS.426..679G,2020MNRAS.499.5486A,2022ApJ...928...10G,2023JCAP...05..033R,2023JCAP...10..016R,2023arXiv231213199R}. Approaches involving more physical models include semi-analytical models \citep{2014MNRAS.443..799O, 2018MNRAS.474..177M,2018MNRAS.481.4221S,2019A&A...631A..82I, 2019MNRAS.490.3667Z,2020MNRAS.497.5432F, 2022MNRAS.510.1880B,2022MNRAS.510.5392K} and hydrodynamical simulations \citep{2015MNRAS.454..269P, 2016MNRAS.461.3563S,2022MNRAS.512..348K,2022MNRAS.512.3243S,2022MNRAS.513.2904T,2023MNRAS.519.1771O, 2023MNRAS.525.5989Y, 2023MNRAS.526.3610H}. 

Numerical simulations \citep{1999MNRAS.303..188K,2001MNRAS.328..726S,2005Natur.435..629S,2011MNRAS.413..101G,2015MNRAS.451.2663H,2015MNRAS.446..521S,2018MNRAS.475..648P} have been successful in investigating contemporary galaxy formation and cosmology. These simulations have shown the ability to reproduce various observed galaxy characteristics across different periods in cosmic history. Their capability in predicting, interpreting, and optimizing observational outcomes has rendered them valuable tools for comprehending the processes underlying galaxy formation and the progression of large-scale structures.
Hydrodynamic cosmological simulations such as EAGLE \citep{2015MNRAS.446..521S} and IllustrisTNG \citep{2018MNRAS.475..648P} have shown enhanced capacity in studying baryonic physics processes, especially gas dynamics. Recently, \citet{2023MNRAS.526.3610H} computed optical and NUV emission lines originating from various sources such as star clusters, narrow-line regions of AGN, post-asymptotic giant branch stars, and fast radiative shocks for galaxies in the IllustrisTNG simulation up to redshift 7 via post-processing, providing valuable predictions for JWST. Similarly, \citet{2023MNRAS.519.1771O} constructed mock H$\alpha$ and [OII] catalogues based on IllustrisTNG to investigate the clustering of emission line galaxies. 
However, such approaches are hindered by substantial CPU time requirements and computational constraints, precluding the simultaneous achievement of both high precision and large simulation volumes.

By combining physically motivated recipes of galaxy formation with N-body cosmological dark matter simulations, semi-analytical models \citep[e.g.][]{1999MNRAS.303..188K,2000MNRAS.319..168C,2011MNRAS.413..101G,2016MNRAS.462.3854L,2018MNRAS.481.3573L} provide a cost-effective alternative. This approach allows simultaneous consideration of high precision and large simulation volumes, as compared to hydro simulations.  \citet{2019A&A...631A..82I} combined the semi-analytical model \lgalaxies{} \citep{2015MNRAS.451.2663H} with the emission line model from \citet{2014MNRAS.443..799O} to construct a light cone for the J-PLUS survey. 
 \citet{2022MNRAS.510.1880B} applied the pre-computed grid of emission line luminosity released by \citet{2016MNRAS.462.1757G} within the semi-analytical galaxy formation code GALFORM \citep{2016MNRAS.462.3854L} to reproduce the observed locus of star-forming galaxies on standard line ratio diagnostic diagrams. \citet{2020MNRAS.497.5432F} applied the emission line model described by \citet{2014MNRAS.443..799O} to three different semi-analytical models: SAG \citep{2018MNRAS.479....2C}, SAGE \citep{2016ApJS..222...22C}, and GALACTICUS \citep{2012NewA...17..175B} and concluded that utilising average star formation rates is a feasible method to generate [OII] luminosity functions. However, these works used different stellar population synthesis (SPS) models for computing stellar components and HII regions, introducing additional inconsistencies in the final results. 

In this study, we implement the state-of-art semi-analytic model \lgalaxies{} \citep{2015MNRAS.451.2663H} onto the merger trees extracted from a large-box-size, high-resolution N-body dark matter simulation to produce a galaxy catalogue for upcoming large-scale surveys. We improve the satellite disruption model in \lgalaxies{} to address a theoretical issue with varying time resolutions. We record the complete star formation history (SFH) for each individual galaxy, enabling the computation of photometric magnitudes for any given filters as post-processes. We combine a grid of HII region models with the public radiation transfer code CLOUDY to derive emission line ratios using the same SPS model employed for the stellar components, ensuring the self-consistency of our predictions.This guarantees consistent treatment between the stellar SED and the emission line luminosities. The grid of HII regions cover a wider parameter space compared to many previous work. By combining this with the semi-analytical galaxy output, we calculate the luminosities of the 13 most frequently utilised NUV and optical emission lines.

This paper is organised as follows. Sec.~\ref{sec:method} provides an overview of the N-body dark matter simulations Jiutian-1G and Mini-Hyper and the details of our semi-analytic model and emission line models. Sec.~\ref{sec:galpred} presents our model predictions for various galaxy properties, while Sec.~\ref{sec:emipred} shows the properties of emission line galaxies. We conclude with a summary and discussion in Sec.~\ref{sec:conc}. 

\section{Data and Method}
\label{sec:method}

In this section, we give a brief description about the dark matter merger trees, semi-analytic models, and emission line models. Two sets of N-body cosmological simulations are adopted, a large simulation, JiuTian-1G, and four sets of merger trees exacted from a small run with different numbers of snapshots. Details are listed in Table~\ref{tab:simulations}. Then we modify the \lgalaxies{} model \citep{2015MNRAS.451.2663H} and apply it on the Jiutian-1G dark matter simulation. We then utilise a publicly available radiation transfer code to determine the luminosity of emission lines by implementing a photoionisation model surrounding the star formation regions.
\begin{table}
	\centering
	\caption{Details about our simulation suite. The first column shows the name of the simulation or merger tree set; the second column shows the corresponding boxsize; the third column shows the particle number; the fourth column shows the dark matter particle mass; the fifth column shows total snapshots.}
	\label{tab:simulations}
	\begin{tabular}{lccccr} 
		\hline
             name & L(cMpc/h) & N & $M_{\rm dm}[\Msunh]$ &  Snapshots\\
		\hline
		JiuTian-1G & 1000 & $6144^3$ & $3.72\times 10^{8}$ & 128\\
		Mini-Hyper-33 & 125 & $768^3$ & $3.67\times 10^{8}$ & 33\\
		Mini-Hyper-65 & 125 & $768^3$ & $3.67\times 10^{8}$ & 65\\
            Mini-Hyper-129 & 125 & $768^3$ & $3.67\times 10^{8}$ & 129\\
            Mini-Hyper-257 & 125 & $768^3$ & $3.67\times 10^{8}$ & 257\\
            
		\hline
	\end{tabular}

\end{table}

\subsection{Dark matter simulation}

\label{sec:dm} 

JiuTian Simulations comprise a series of cosmological N-body simulations, ranging in box size and resolution. The JiuTian-1G (hereafter JT1G) simulation is a large dark matter only $N$-body simulation within the framework of $\Lambda$ cold dark matter ($\Lambda$CDM) cosmology designed for next-generation surveys. Utilising the L-Gadget3 code \citep{2005MNRAS.364.1105S}, the JT1G simulation tracks $6144^3$ dark matter particles within a cubic simulation box with a side length $L_{\rm box}$ = 1Gpc/h. This box length is twice as large as the previous Millennium Simulation \citep[MS,][]{2005Natur.435..629S}, resulting in an eight-fold increase in volume compared to MS. The particle mass is $3.72 \times 10^{8} \Msunh$, almost three times smaller than the original MS. 
The JT1G simulation stores 128 snapshots ranging from redshift 127 to 0, with an average time gap of approximately 100 Myr. This time resolution is chosen for weak lensing studies and is twice as large as MS with 64 snapshots. We adopt the cosmological parameters from \citet{2020A&A...641A...1P} as follows: $\sigma_8=0.8102$, $H_0=67.66{\rm km s^{-1} Mpc^{-1}}$, $\Omega_\Lambda=0.6889$, $\Omega_{\rm m}=0.3111$, $\Omega_{\rm b}=0.0490 (f_{\rm b}=0.1575)$. 
 Following \citet{2005Natur.435..629S}, dark matter halos and subhalos are identified using the friends-of-friends (FOF) and SUBFIND \citep{2001MNRAS.328..726S} algorithms. Additionally, to establish the merger trees, the subhalos are linked with their unique descendants employing the B-Tree code. Details about JT1G are referred to Han et al. in prep and Li et al. in prep. 

Although previous works have extensively examined the particle mass resolution \citep[see][for a review]{2023ARA&A..61..473C}, limited attention has been given to time resolution. \citet{2012MNRAS.419.3590B} concluded that 128 snapshots are necessary for the GALACTICUS semi-analytical model \citep{2012NewA...17..175B} to achieve convergence within a 5\% level in stellar mass. We conduct four sets of merger trees with different temporal resolutions from a smaller simulation, Mini-Hyper, to assess the time convergence. The parent simulation has a box size of 125\mpch{} and a particle mass similar to JT1G, $3.674 \times 10^{8} \Msunh$. The total particle number is $768^3$ which is stored in 513 distinct snapshots. We employ slightly different cosmological parameters compared to JT1G: $\sigma_8=0.826$, $H_0=67.3{\rm km s^{-1} Mpc^{-1}}$, $\Omega_\Lambda=0.693$, $\Omega_{\rm m}=0.307$, $\Omega_{\rm b}=0.04825 (f_{\rm b}=0.1572)$. 

\begin{figure}
\centering
\includegraphics[width=\columnwidth]{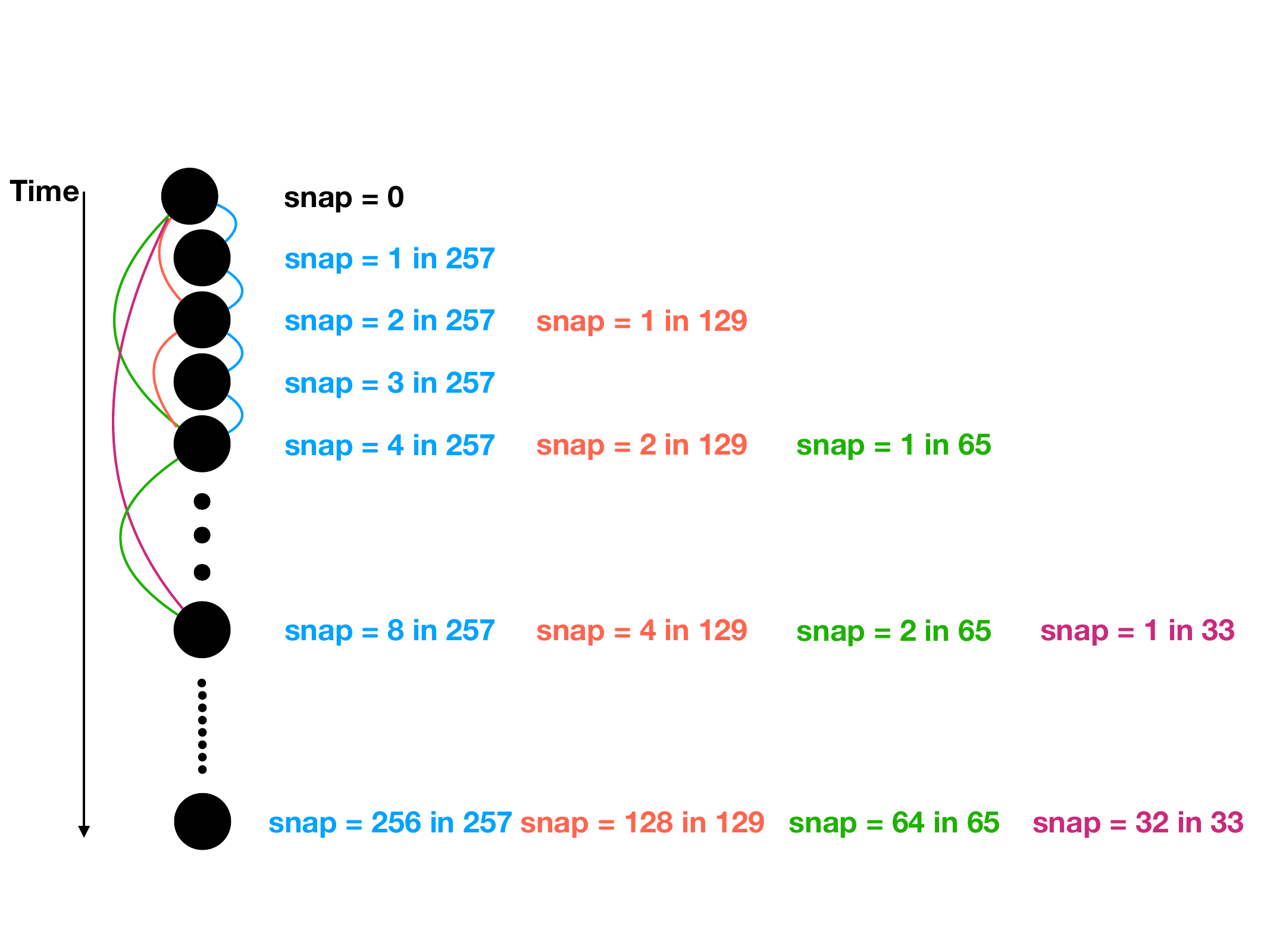}
\caption{An example of constructing merger trees with varying time intervals: using the whole 257 snapshots, we create the Mini-Hyper-257, represented by the blue lines. By skipping half of the snapshots, we construct the Mini-Hyper-129, shown by the red lines. Continuing this pattern, we skip half of the Mini-Hyper-129 snapshots to generate the Mini-Hyper-65 (green lines). We employ a similar methodology to generate merger trees with fewer snapshots, Mini-Hyper-33(purple lines).}
\label{fig:mergertrees}
\end{figure}

Four merger trees are then constructed accordingly with varying time intervals using the B-Tree code with different "SnapSkipFac". This approach yields four sets of merger trees with comparable merger tree structures but differing numbers of snapshots. Table~\ref{tab:simulations} shows the parameters for all our simulations. In practice, we fix the first and last snapshots across all simulations. Then we select every second snapshot to construct a simulation with 257 snapshots. Further skipping every other snapshot results in a simulation with 129 snapshots. Following the same methodology, simulations with 65/33 snapshots were constructed, with the number of snapshots decreasing by a factor of 2 each time. Therefore, we obtain four sets of merger trees with similar tree structures, wherein the time intervals between two snapshots increase by a factor of 2 each time. It is worth noting that these four sets of merger trees have the same dark matter halo properties at common redshifts. Fig.~\ref{fig:mergertrees} depicts how we create merger trees with different snapshots. 

\subsection{Semi-Analytical Model: L-Galaxies}
\label{sec:sam}

We use the version of the \lgalaxies{} code as described in \citet{2015MNRAS.451.2663H} (hereafter H15) and make modifications to solve the time convergence problem. H15 includes physical prescriptions for various baryonic processes, such as shock heating, gas cooling, star formation, supernova feedback, formation and growth of supermassive black holes (SMBH), AGN feedback, metal enrichment and etc. Details about the physical recipes and parameters can be found in their supplementary material. We have modified the satellite disruption procedure to address the issue of time convergence and have readjusted the parameters to replicate the abundance of SMBH in the local Universe.
\subsubsection{Time convergence problem}

\begin{figure*}
\centering
\includegraphics[width=1.5\columnwidth]{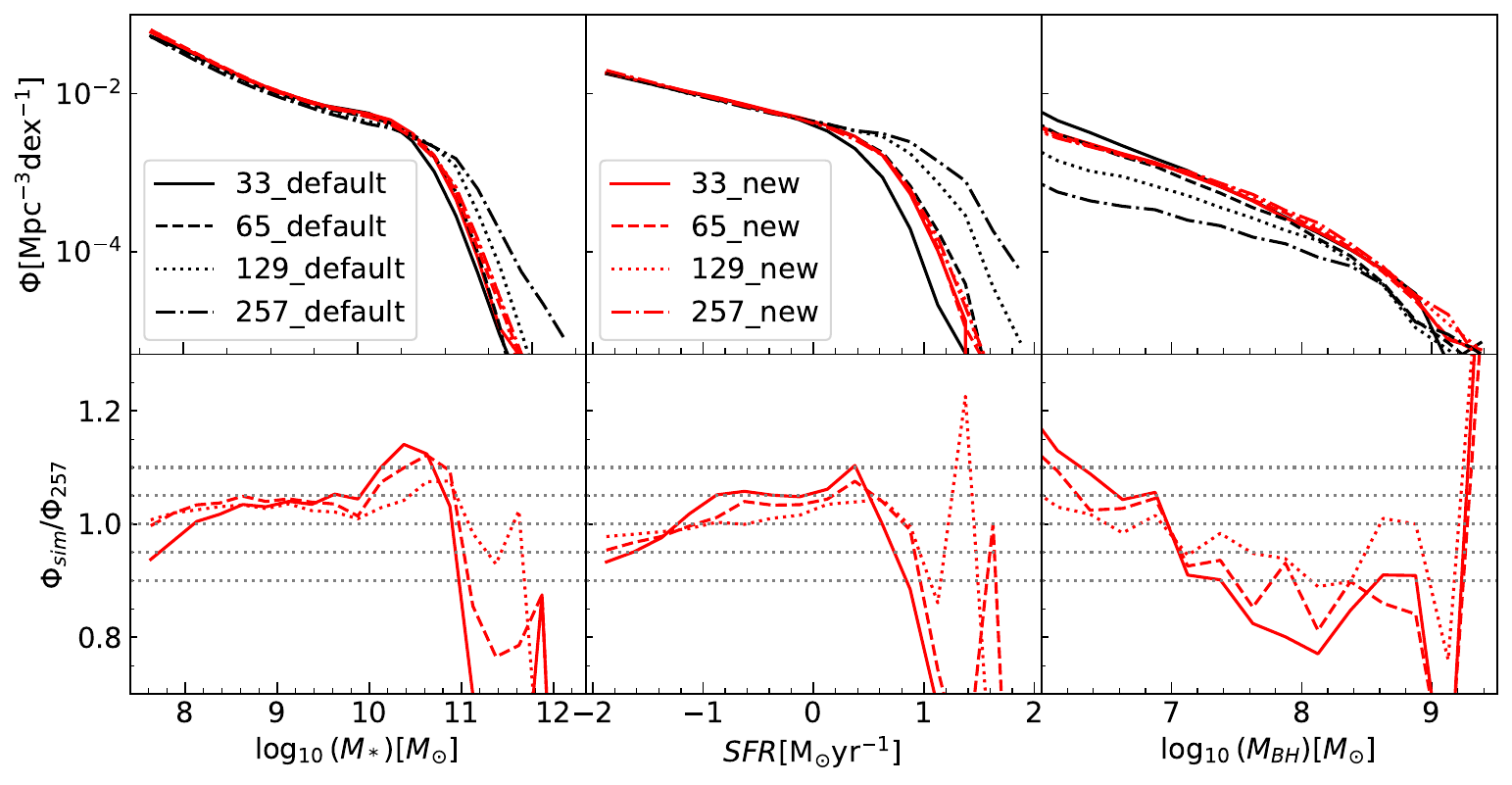}
\caption{Galaxy properties predicted by the original H15 model and modified model in Mini-Hyper simulations with different snapshots. The black lines represent the H15 results, while the red lines represent results from our modified model. Solid lines correspond to Mini-Hyper-33, dashed lines to Mini-Hyper-65, dotted lines to Mini-Hyper-129, and dotted-dashed lines to Mini-Hyper-257. Left panel: Stellar Mass Function (SMF); middle panel: Star Formation Rate Function (SFRF); right panel: Black Hole Mass Function (BHMF). The bottom row quantifies the differences normalised to the Mini-Hyper-257 simulation.}
\label{fig:conv}
\end{figure*}

We utilise the original H15 code to examine whether similar galaxy properties can be acquired on dark matter merger trees with varying time intervals. 
 The black lines in the first row of Fig.~\ref{fig:conv} show the statistical properties of galaxies at z$\sim$0 predicted by the original H15 model, including the galaxy stellar mass function (SMF), galaxy abundance as a function of star formation rate (SFRF) and the supermassive black hole mass function (BHMF). Distinct line styles represent results from merger trees with different time intervals: solid lines for Mini-Hyper-33, dashed lines for Mini-Hyper-65, dotted lines for Mini-Hyper-129, and dotted-dashed lines for Mini-Hyper-257 merger trees. Surprisingly, we observe substantial differences in galaxy properties across simulations. The left panel reveals that simulations with smaller time intervals tend to have more massive galaxies. The difference is remarkably large, reaching several orders of magnitude at $M_*>10^{11}\Msun$ between Mini-Hyper-33 and Mini-Hyper-257. Larger offsets in SFRF and BHMF are evident, as the Mini-Hyper-257 showcases significantly higher numbers of highly star-forming galaxies and considerably less SMBH compared to the other simulations. These substantial differences with different time intervals strongly suggest a challenge in time convergence within the H15 code.

\subsubsection{Relevant physical processes}
\label{sec:rpp}

Further investigation into the H15 code reveals that the primary cause of the time convergence problem is linked to the fate of orphan satellite galaxies, which lost their subhalos due to physical processes or numerical effects. 
In H15, following the disruption of its subhalo, a merging clock is set simultaneously based on an estimated time ($t_{\rm friction}$) for the orphan to spiral into the central galaxy due to dynamical friction. 
An orphan galaxy will ultimately either undergo disruption or merge. It could suffer tidal disruption on its way spiralling into the centre, relying on the competition between self-gravity and the tidal force from the main dark matter halo. In practice, a comparison is made between the baryonic (cold gas and stellar mass) density within the half-mass radius and the dark matter density of the main halo at the assumed pericentre ($R_{\rm peri}$) of the orphan's orbit. 

If the tidal force exceeds the bounding gravity, prior to $\Delta t = t_{\rm friction}$ where $\Delta t$ is the time since merger clock is set, the orphan galaxy will be completely disrupted and no longer undergo merging. Their SMBH, gas, and stars are all distributed in the halo of the central galaxy. SMBH in the central galaxy will not grow in this scenario. Conversely, if tidal force never exceeds the bounding gravity (until $\Delta t = t_{\rm friction}$), the orphan galaxy will merge into the central galaxy at $\Delta t = t_{\rm friction}$. During galaxy mergers, the SMBH in the central galaxy could grow by swallowing the SMBH from the satellite galaxies that are being merged and by experiencing strong gas accretion, which is triggered by the merger process.

A residual time-dependent treatment was adopted in the original H15 model, where the merger recipe and the satellite disruption recipe are called at different times. H15 divides the time between two adjacent snapshots into 20 sub-steps and calls the "merger" recipe in each sub-step, while the "disruption" recipe is only called at the end of each snap gap. This prioritises the occurrence of merger, while the disruption is delayed until the final sub-step, regardless of meeting the disruption criterion earlier. Larger intervals of time between two consecutive snapshots increase the chances of mergers occurring. As a result, there are fewer disrupted orphan galaxies, and the SMBH become more massive. The increased mass of SMBH in turn leads to more effective active galactic nucleus (AGN) feedback, resulting in smaller central galaxies. We conducted experiments to preserve time resolution in the substep by adjusting the number of substeps in different Mini-Hyper simulations, but this did not solve the time resolution problem.

\subsubsection{Modifications to the disruption model}

To address this time convergence problem, we make modifications to the disruption model in the code. We replace the assumed peri-centre distance with the actual distance of orphan satellite galaxies from the central galaxy $R_{\rm orphan}$, and call the disruption function at each substep. We track the most bound particle, which was defined at the point just before the orphan galaxy lost its substructure. By multiplying its distance from the central galaxy $R_{\rm mostbound}$ by a factor that accounts for the impact of dynamical friction, we can estimate the distance of the orphan galaxy $R_{\rm orphan}$ as follows.
\begin{equation}
    R_{\rm orphan} = \sqrt{1-\frac{\Delta t}{t_{\rm friction}}}R_{\rm mostbound}.
\end{equation}

We introduce a minimum distance at which disruption could occur, set as the scale radius of the gas disk in the central galaxy, $R_{\rm gas}$. The choice of $R_{\rm gas}$ as the minimum distance is justified by the notion that if an orphan galaxy has reached the region of $R_{\rm gas}$, it should be considered as having entered the region of central galaxies and is undergoing a strong interaction. In such situations, it is more appropriate to treat the event as a merger rather than a disruption. 

We implement the modified model to the Mini-Hyper simulations with different snapshots without making any changes to the parameters from the initial H15 version.
The stellar mass function, star formation rate function, and black hole mass function of the modified model are shown as the red lines in the first row of Fig.~\ref{fig:conv}, indicating good agreement among different time resolutions. The quantification of the differences with respect to the Mini-Hyper-257 simulation is presented in the second row and shows that the differences in SMF and SFRF are within 5\% in most cases. At higher values, it could suffer from the limited sample size. The variation in BHMF seems larger, especially at the low mass and at the high mass end. This suggests that the growth of SMBH could be more sensitive to the time interval between consecutive snapshots. Furthermore, varying numbers of snapshots could also result in difference within the generated trees. For example, \citet{2016MNRAS.459.1554W} showed that using more snapshots typically results in shorter branches for most tree builders. This is because more linking errors may occur when processing more snapshots, as tree builders are prone to resolution or flip-flop problems. \citet{2018MNRAS.474..604H} also showed that the \textsc{SUBFIND} and \textsc{DTree} combination could produce a substantial amount of fragmented branches, which in turn impacts the properties of resulting galaxies.

\subsubsection{Parameter adjustment}
\label{sec:para}
It has been noticed that the mass of SMBH is lower by an order of magnitude compared to current observations \citep[e.g.][]{2019MNRAS.483..503Y} at z = 0 (see also the upper panel in Fig.~\ref{fig:bh}, the black line is the result of H15, the purple symbols are observational data). To determine the parameters of our new model, we incorporate the black hole mass function at z = 0 as constraint, in addition to the observed galaxy stellar mass function at z=0, 1, 2, and 3, and the passive fraction at z = 0.4. 
The observational data we use in this study are listed in Table~\ref{tab:obs}.
In accordance with \citet{2013MNRAS.431.3373H}, we establish a representative subset of subhalo merger trees and employ the MCMC method as described in \citet{2009MNRAS.396..535H}, to thoroughly explore the multidimensional parameter space with the updated disruption model in \lgalaxies{}. 
In short, we divide the haloes into \textit{I} halo mass bins with a width of 0.5 dex, and the galaxies into \textit{J} stellar mass bins with a same width. We randomly select $n_i$ haloes from a total $N_i$ halos in the $i$th halo mass bin. The number $n_i$ is determined by a set of linear constraint equations, $\sum_{i=1}^I \frac{N_i^2}{n_i^2} \phi_{i j}=F^2 \Phi_j^2$, where $\phi_{i j}$ is the average number of galaxies in the $j$th stellar mass bin for haloes in the $i$th halo mass bin, $\Phi_j$ is the total number of galaxies in the $j$th stellar mass bin. $F$ is the uncertainty of the stellar mass function, which we set to be $F<0.05$. Our final representative subsample is $\sim$ 1/512 of the whole box. Details about the sample construction is available in \citet{2013MNRAS.431.3373H}, APPENDIX B.

For comparison with the default H15 parameters, the best-fit parameters are enumerated in Table~\ref{tab:para}.
Our modified code maintains compatibility with the default H15 parameters. 
The best-fit parameters are then applied to the full volumes of the JT1G simulation to generate the SAM galaxy catalogue.

\begin{table*}
\begin{center}
  \caption{Results from the MCMC parameter estimation. The best-fit
    values of parameters are compared with the values published in \citet{2015MNRAS.451.2663H}.}
\label{tab:para}
\begin{tabular}{lccc}
\hline
\hline
&JT1G &Henriques15 &Units\\
\hline
\hline
$\alpha_{\rm{SF}}$ (SF eff) &0.033 &0.025\\
$\Sigma_{\rm{SF}}$ (Gas density threshold) &0.57 &0.24 &$\rm{10^{10}\Msun\,pc^{-2}}$\\
$\alpha_{\rm{SF, burst}}$ (SF burst eff) &0.135 &0.60 \\
$\beta_{\rm{SF, burst}}$ (SF burst slope) &0.70 &1.9 \\
\hline
\hline
$k_{\rm{AGN}}$ (Radio feedback eff) &$5.76\times 10^{-4}$ &$5.3\times10^{-3}$ & $\rm{\Msun\,yr^{-1}}$\\
$f_{\rm{BH}}$ (BH growth eff) &0.0045 &0.041 \\ 
$V_{\rm{BH}}$ (Quasar growth scale) &230 &750 &$\rm{km\,s^{-1}}$\\ 
\hline
$\epsilon$ (Mass-loading eff) &2.78 &2.6  \\
$V_{\rm{reheat}}$ (Mass-loading scale) &924 &480  &$\rm{km\,s^{-1}}$\\
$\beta_{1}$ (Mass-loading slope) &0.124 &0.72  \\
$\eta$ (SN ejection eff) &2.16 &0.62 \\
$V_{\rm{eject}}$ (SN ejection scale) &250 &100  &$\rm{km\,s^{-1}}$\\
$\beta_{2}$ (SN ejection slope) &0.197 &0.80 \\
$\gamma$ (Ejecta reincorporation)
&$1.5\times10^{10}$ &$3.0\times10^{10}$   & yr\\
\hline
$M_{\rm{r.p.}}$ (Ram-pressure threshold) &$1.5\times10^{4}$ &$1.2\times10^{4}$ &$\rm{10^{10}\Msun}$\\
$R_{\rm{merger}}$ (Major-merger threshold) &0.1 &0.1 \\
$\alpha_{\rm{friction}}$ (Dynamical friction) &4.0 &2.5 \\
\hline
$y$ (Metal yield) &0.050 &0.046 \\
\hline
\hline
\end{tabular}
\end{center}
\end{table*}

\subsubsection{Mass-resolution convergence test}
We have evaluated the mass resolution effect by applying the same SAM models (our modified model) and parameters (the best-fitting parameters for JT1G) to the merger trees extracted from the re-scaled Millennium-II simulation \citep{2009MNRAS.398.1150B,2015MNRAS.451.2663H}. The MSII simulation has a mass resolution of $7.69\times 10^{6}\Msunh$, which is $\sim$ 50 times higher than JT1G, although the volume is smaller by 1000 \citep[see][and reference therein for more details about the re-scaled MSII]{2015MNRAS.451.2663H}. MSII has been shown to resolve the smallest halo capable of hosting a detectable galaxy \citep[e.g. ][]{2011MNRAS.413..101G}. Fig.~\ref{fig:conv_MSII_SMF} illustrates that the stellar mass of the galaxy converges at $10^{9}\Msun$ between JT1G and MSII within a 10\% from z=0 - 3. This difference increases to 30\% at $10^{8}\Msun$, partly attributed to slight differences in the cosmological parameters and possibly stemming from distinctions in their initial conditions (Li et al. in prep). 
At high masses, the disparity between JT1G and MSII is mainly due to varying mass resolutions. The mass resolution in MRII is about 50 times higher than in JT1G, allowing it to resolve more small halos that can later merge into larger systems. These mergers lead to larger supermassive black holes (SMBHs) in MRII, resulting in more effective AGN feedback that suppresses star formation and ultimately leads to a smaller stellar mass. The difference in stellar mass in massive halos is not significant, typically around 0.15 dex. However, due to the steep slope at the high mass end of the SMF, this slight variation in stellar mass can result in a notable difference in the stellar mass function at high masses. Along with mass resolution, cosmic variance stemming from the smaller volume of MSII could also be a factor. Similar comparisons are performed on the abundance of the galaxy as a function of SFR in Fig.~\ref{fig:conv_MSII_SFRF}. It shows a very good convergence between the JT1G simulation and MSII at z =0. At higher redshifts, the convergence is somehow larger, but all within 10\%.

\subsection{Emission line model}
\label{sec:elm}
In this section, we first briefly describe the process of generating the galaxy spectral energy distribution (SED). Then, we explain in detail how to use the expected SED as the input ionising spectrum to calculate the luminosities of emission lines. In practice, we employ the radiative transfer code CLOUDY to calculate the relative strength of emission lines on a grid of HII region models. Meanwhile, we adopt empirical relations to establish connections between the general properties of galaxies and the parameters that describe the ionisation regions. According to the general properties of a given galaxy, the luminosity of each emission line is derived by performing interpolation within a pre-computed grid of line luminosities. During this process, we utilise the same Stellar Population Synthesis model and initial mass function as those employed in calculating the galaxy SED, ensuring the self-consistency of galaxy SED and emission line calculation. 

\subsubsection{Galaxy spectral energy distributions}
\label{sec:sed}

Stellar Population Synthesis (SPS) models serve as essential tools in astrophysics, enabling the generation of synthetic SEDs with detailed information about the star formation history (SFH), metallicity, and initial mass function (IMF). These models provide valuable insights into the formation and evolution of galaxies. 
In this work, our default SPS model is based on the \citet{2003MNRAS.344.1000B} framework, with a Chabrier IMF \citep{2003PASP..115..763C}. 

The entire SFH of both the disk and bulge components is stored in 22 distinct bins, as detailed by \citet{2015MNRAS.451.2681S}. Consequently, the SED can be produced using any desired SPS model and IMF as post-processes.

\subsubsection{Grid of emission line models}
\label{sec:lineratio}
We use the c17.04 release of the photoionisation code CLOUDY \citep{2017RMxAA..53..385F}, a widely used tool, to address radiation transfer in photoionised regions. To solve the radiation transfer equation, we specify the most important input physical properties of the gas cloud and spectrum of ionising sources, including the intensity and spectrum of the ionising photons, gas geometry, gas metallicity $Z$, and hydrogen density $n_{\rm H}$. 

 We adopt the BC03 \citep{2003MNRAS.344.1000B} SPS model with a Chabrier IMF \citep{2003PASP..115..763C} to calculate the intensity and spectrum of ionising photons. These SPS model and IMF are also used in calculating the galaxy SED. It is noteworthy that previous studies did not consistently employ the same stellar population models for computing the galaxy SED and for the photoionisation modelling of nebular emission. For example, \citet{2022MNRAS.510.1880B} used the M05 model \citep{2005MNRAS.362..799M} to calculate galaxy stellar SED, yet adopted the emission line models based on BC03 in the HII regions.  \citet{2019A&A...631A..82I} adopted different stellar synthesis models to calculate the emission line grid \citep{2010AJ....139..712L} and the stellar SED \citep{2003MNRAS.344.1000B} when generating the emission line galaxy mock catalogue for J-PLUS. This difference results in a lack of self-consistency between the two distinct components of the combined galaxy spectrum (i.e. stellar and emission). In contrast, our work ensures internal consistency by employing the same SPS model (BC03) and IMF (Chabrier) for both the stellar component and the nebular model within galaxies.

For the gas geometry, we follow \citet{2017ApJ...840...44B} to assume a spherical shell geometry with the ionising source located in the centre, and to adopt an inner radius $R_{\rm inner}$ at $10^{19}$ cm. 

In contrast to \citet{2017ApJ...840...44B}, who assumes a constant gas density, we take into account the gas density within various ranges. In practice, we calculate the emission line ratios for a grid of HII region models by sampling $\mathcal{U}$, $Z$, and $n_{\rm H}$ as follows:
$$
\begin{aligned}
& \log _{10} \mathcal{U}:-4.0,-3.5,-3.0,-2.5,-2.0,-1.5,-1.0 \\
& \log _{10} Z / Z_{\odot}:-2.0,-1.5,-1.0,-0.6,-0.4,-0.3,-0.2, \\
& -0.1,0.0,0.1,0.2 \\
& \log _{10} n_{\rm H}: 1,2,3,4\  \text{(}\mathrm{cm}^{-3}\text{)}
\end{aligned}
$$
where $\mathcal{U}$ is the ionisation parameter, a dimensionless parameter defined as the ratio of hydrogen ionising photons to the total hydrogen density:
\begin{equation}
\label{U}
    \mathcal{U} \equiv \frac{n_{\gamma}}{n_{\rm H}},
\end{equation}
where $n_{\gamma}$ is the volume density of the ionizing photons. $Z$ is the metallicity of cold gas, and $n_{\rm H}$ is the hydrogen density.
The output is a grid of line ratios relative to the H${\alpha}$ luminosity with different parameters. Using linear interpolation in $\log _{10} \mathcal{U}$, $\log _{10} Z / Z_{\odot}$ and $\log _{10} n_{\rm H}$, we extract line ratios for each star-forming galaxy. If the values for $\mathcal{U}$, $Z$, and $n_{\rm H}$ fall outside the range of the grid, we use the closest limiting values from the grid to align with the prediction of the catalogue. 

\begin{figure}
\includegraphics[width=\columnwidth]{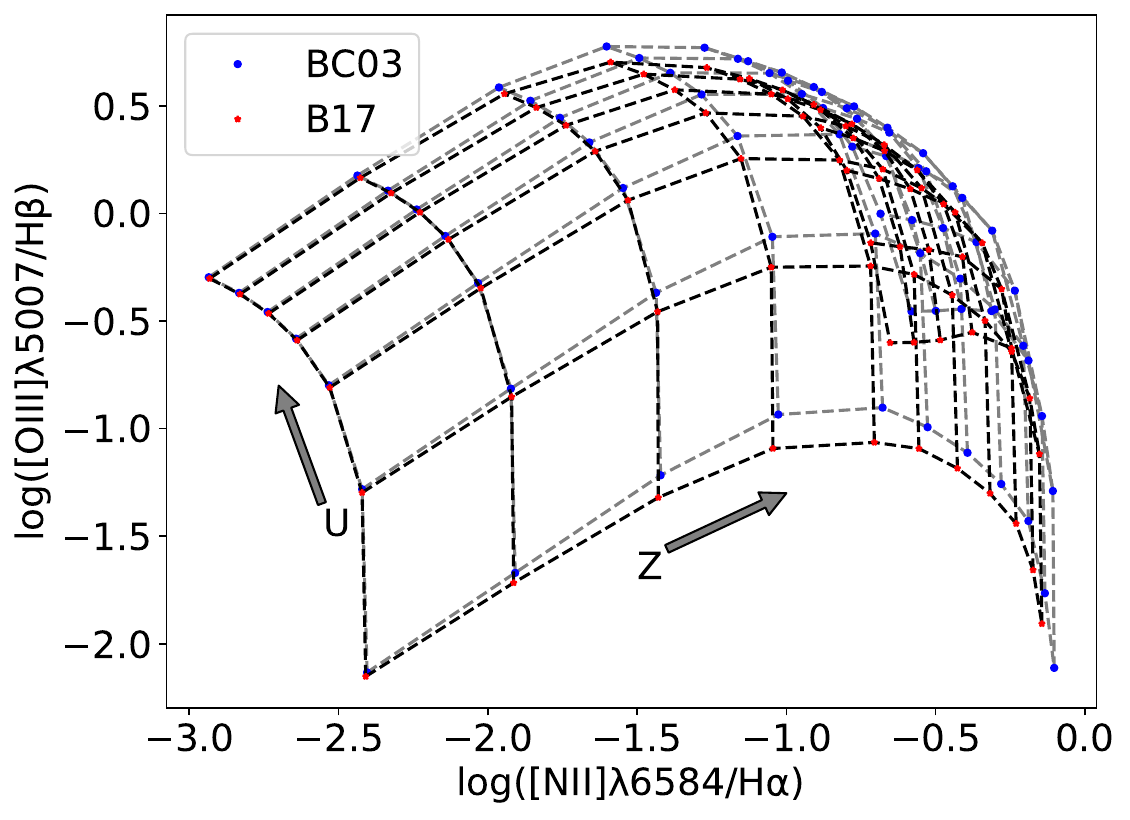}
\caption{The classic BPT diagram illustrating the pre-computed grid of emission line luminosities. The x-axis represents ${\rm [NII]\lambda 6584/H\alpha}$, while the y-axis represents ${\rm [OIII]\lambda 5007/H\beta}$. The gray grid with blue symbols denote the grid corresponding to a hydrogen density of $n_{\rm H} = 10 {\rm cm^{-3}}$, with variations in metallicity $Z$ and ionisation parameter $\mathcal{U}$. Additionally, the black lines with red symbols depict the emission line ratio grid from \citet{2017ApJ...840...44B}.}
\label{fig:BPT}
\end{figure}

 Fig.~\ref{fig:BPT} shows the classic BPT diagram \citep{1981PASP...93....5B} of the pre-computed grid of emission line luminosities. The x-axis depicts ${\rm [NII]\lambda 6584/H\alpha}$, and the y-axis depicts ${\rm [OIII]\lambda 5007/H\beta}$. The gray grid with blue symbols shows the grid with a hydrogen density of $n_{\rm H} = 10 {\rm cm^{-3}}$, varying the metallicity $Z$ and the ionisation parameter $\mathcal{U}$. Additionally, we present the grid of emission line ratios from \citet{2017ApJ...840...44B} as black lines with red symbols. We employ a hydrogen density, metallicity, and ionisation parameter grid akin to that utilised in \citet{2017ApJ...840...44B}, albeit with a different spectrum of ionising photons. Furthermore, we use the Chabrier IMF, while they employ the Kroupa IMF \citep{2001MNRAS.322..231K}. Our BPT diagram bears a resemblance to theirs, with a slight offset observed at higher metallicity. The ${\rm [OIII]\lambda 5007/H\beta}$ ratio exhibits a monotonically increasing trend with increasing $\mathcal{U}$, while the ${\rm [NII]\lambda 6584/H\alpha}$ ratio demonstrates a monotonically increasing pattern with increasing $Z$. We notice that as $Z$ becomes sufficiently large, an overlapping within the grid itself becomes evident, reflecting the degeneracy of $Z$ and $\mathcal{U}$.

\subsubsection{H${\alpha}$ luminosity}
CLOUDY provides line ratios relative to the H${\alpha}$ luminosity, which is closely related to the star formation rate. For a nebula that is absolutely optically thick for ionising photons and optically thin for redward photons \citep[case B,][]{2006agna.book.....O}, we can theoretically derive the relation between the intensity of a specific hydrogen recombination line and the ionising photon rate by quantum mechanics. The relation between the luminosity of H${\alpha}$ and the ionising photon rate is expressed as:
\begin{equation}
\label{eq:halpha}
    L(\mathrm{H} \alpha)=\frac{\alpha_{\mathrm{H} \alpha}^{\mathrm{eff}}}{\alpha_{\mathrm{B}}} h \nu_{\mathrm{H} \alpha} Q_{\mathrm{H}}=1.37 \cdot 10^{-12} Q_{\mathrm{H}},
\end{equation}
where $\alpha_{\mathrm{H} \alpha}^{\mathrm{eff}}$ is the effective recombination coefficient at H${\alpha}$, and $\alpha_{\mathrm{B}}$ is the case B recombination coefficient. $Q_{\mathrm{H}}$ is the total number of ionised photons emitted per second,
which could be related to the star formation rate assuming the Chabrier IMF \citep{2003PASP..115..763C}:
\begin{equation}
\label{eq:sfr_q}
    Q_{\rm H} = 1.35 \cdot 10^{53} \operatorname{SFR}.
\end{equation}
The H${\alpha}$ luminosity can then be expressed as a function of SFR \citep{2013seg..book.....F}:
\begin{equation}
\label{eq:ha_sfr}
    L(\mathrm{H} \alpha) = 1.863 \cdot 10^{41} \operatorname{SFR}, 
\end{equation}
It is noteworthy that we use the same IMF in this $L(\mathrm{H} \alpha) - {\rm SFR}$ relation as employed in both the stellar component and the photoionization model, which ensures the overall consistency of our model.

 When paired with the line-ratio produced by CLOUDY as outlined in Sec.~\ref{sec:lineratio}, we are able to calculate the luminosity of any specific emission line at $\lambda_{j}$.
\begin{equation}
\label{eq:lj}
    L\left(\lambda_j\right)=L(\mathrm{H} \alpha)\times R\left(\lambda_j, \mathcal{U}, Z, n_{\rm H}\right),
\end{equation}
where $R\left(\lambda_j, q, Z, n_{\rm H}\right)$ is the CLOUDY predicted ratio of the desired emission line at wavelength $\lambda_j$ to H${\alpha}$ with a given set of $\mathcal{U}$, $Z$ and $n_{\rm H}$. 

 We include the 13 most widely emission lines in NUV and optical ranges in the final catalogue \footnote{The whole line ratio table is available at: https://github.com/peiwenxiang/EmissionLineGrid\_BC03}, as listed in Table ~\ref{tab:emi}. Additional emission lines can be provided upon request.

\begin{table}
	\centering
	\caption{Details of the 13 emission lines provided by our study.}
	\label{tab:emi}
	\begin{tabular}{lcr} 
		\hline
             name & wavelength (\AA) \\
		\hline
            $\rm Ly{\alpha}$ & 1216 \\
            $\rm H{\beta}$ & 4561 \\
            $\rm H{\alpha}$ & 6563 \\
            $\rm [OII]$3727 & 3727 \\
            $\rm [OII]$3729 & 3727 \\
            $\rm [OIII]$4959 & 4959 \\
            $\rm [OIII]$5007 & 5007 \\
            $\rm [OI]$6300 & 6300 \\
            $\rm [NII]$6548 & 6548 \\
            $\rm [NII]$6584 & 6584 \\
            $\rm [SII]$6717 & 6717 \\
            $\rm [SII]$6731 & 6731 \\
            $\rm [NeIII]$3870 & 3870 \\
		\hline
	\end{tabular}
\end{table}

\subsubsection{Emission line luminosity for semi-analytical galaxies}
With the grid of emission lines, the final step in predicting the emission lines for semi-analytical galaxies is to establish a connection between the general properties of the galaxies and the parameters that determine the emission lines. These parameters include gas metallicity, hydrogen density, and ionisation parameters. 

The gas metallicity is obtained directly from the semi-analytical catalogue. Here we use gas metallicity $Z_{\rm cold}$ for the whole galaxy. Calculations including individual heavy elements will be performed in further work.

 We adopt the empirical relations based on local observations \citep{2019MNRAS.486.1053K} to link the hydrogen density $n_{\rm H}$ to the stellar mass and specific star formation rate as follows, 
\begin{equation}
\begin{aligned}
\log _{10}\left[\frac{n_{\mathrm{H}}}{\mathrm{cm}^{-3}}\right] & =2.066+0.310\left(\log _{10}\left(M_* / \Msun \right)-10.0\right) \\
& +0.492\left(\log _{10}\left({\rm s S F R} / \mathrm{yr}^{-1}\right)+9\right) 
\end{aligned}
\end{equation}
where sSFR is the specific star formation and $M_*$ is the stellar mass. Both can be obtained directly from the semi-analytical catalogue.

The joint effect of the ionising spectrum and its intensity can be described by the ionisation parameter, $\mathcal{U}$. 
We adopt the empirical relations based on local observations \citep{2019MNRAS.486.1053K} to link the ionization parameter to the specific star formation rate, gas metallicity and hydrogen density as follows,  
\begin{equation}
\begin{aligned}
\log _{10} \mathcal{U} & =-2.316-0.36\left(0.69+\log _{10}\left(Z_{\mathrm{cold}} / Z_{\odot}\right)\right) \\
& -0.292 \log _{10}\left(n_{\mathrm{H}} / \mathrm{cm}^{-3}\right) \\
& +0.428\left(\log _{10}\left({\rm s S F R} / \mathrm{yr}^{-1}\right)+9\right).
\end{aligned}
\end{equation}
So far, we have obtained the input parameters for the corresponding galaxy. By interpolating within a pre-computed grid of line luminosities, as explained in the previous section, one can determine the luminosity of every emission line for each model galaxy.

\begin{figure*}
\centering
\includegraphics[width=2\columnwidth]{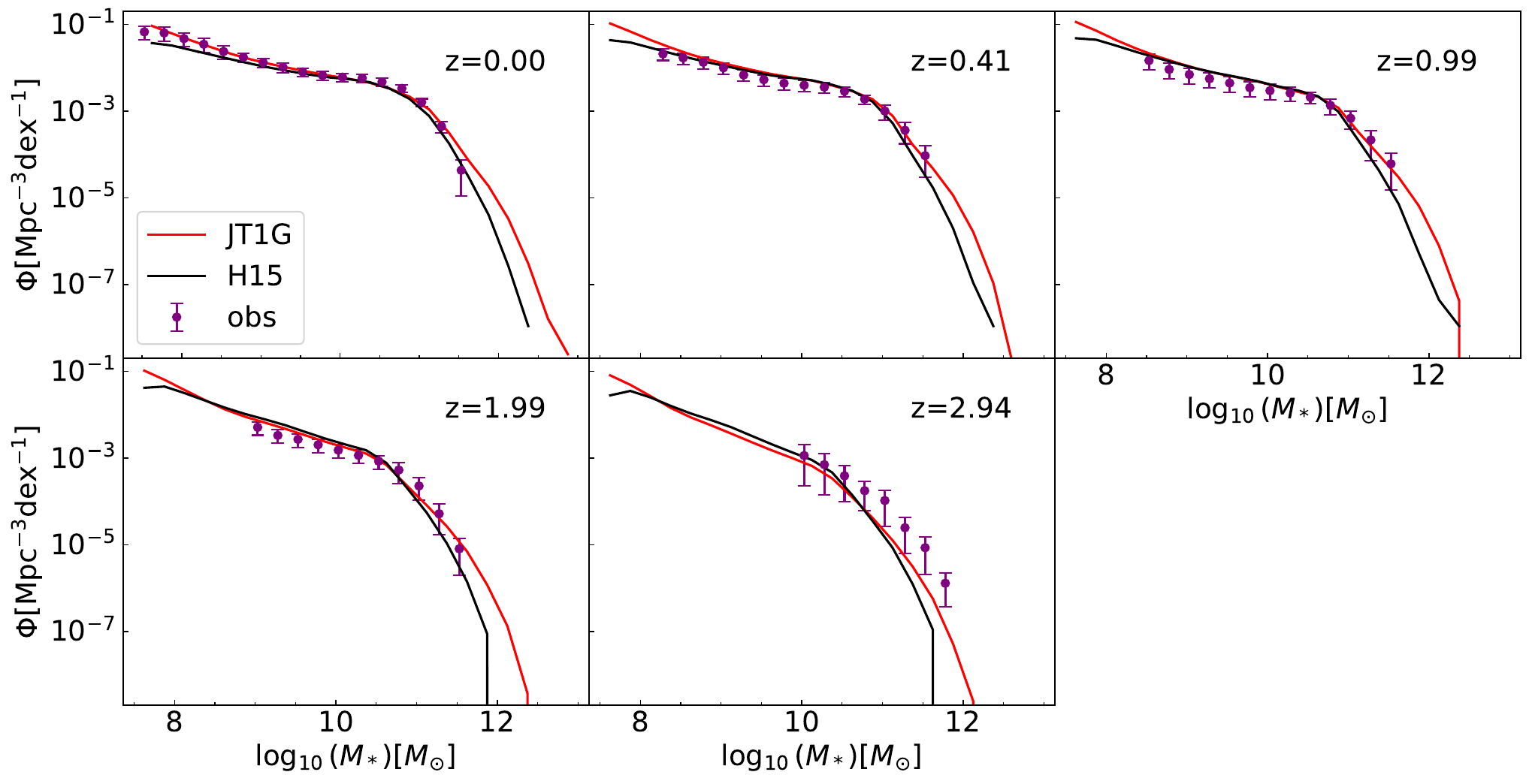}
\caption{Galaxy Stellar Mass Function (SMF) spanning redshift 0 to 3. The red lines represent the results of our model, while the black lines represent the results of MR from H15. Purple symbols with error bars denote observational data points used in our MCMC procedures, combining various studies to estimate systematic uncertainties (see Appendix A of H15 for details). }
\label{fig:smf}
\end{figure*}

The procedure for generating emission lines is generally similar to previous works \citep{2014MNRAS.443..799O, 2019A&A...631A..82I,2020MNRAS.497.5432F} but varies in details. Our pre-calculated emission line models rely on gas density, metallicity, and ionization parameter. These previous works assumed a fixed $n_{\rm H} = 100 \rm cm^{-3}$, with the photoionization parameter depends on gas metallicity. Consequently, their emission line models are solely influenced by metallicity. Furthermore, we employ the same BC03 SPS model and Chabrier IMF for determining both stellar SED and emission line luminosities, thus ensuring internal consistency within the model. \citet{2014MNRAS.443..799O}, \citet{2019A&A...631A..82I}, and \citet{2020MNRAS.497.5432F} utilize the Salpeter IMF for computing emission line ratios but switch to the Kroupa IMF for calculating the H$\alpha$ luminosity. \citet{2022MNRAS.510.1880B} introduced a distinct SPS model for stellar SED (M05) and nebular regions (BC03). The main difference between our approaches and \cite{2018MNRAS.474..177M} lies in the use of different links between galaxy properties and the input parameters for CLOUDY. For example, they uses an average gas density, the total gas mass divided by the cubic radius, whereas we implemented a empirical scaling relation associating the SFR and stellar mass with typical gas density in the SFR vicinity.

\subsection{Dust extinction}
\label{sec:dust}

The extinction induced by dust significantly influences the observed spectra of galaxies, absorbing UV/optical photons and re-emitting them at longer wavelengths. Consequently, galaxies rich in dust tend to have red colours even if they have a high SFR. In this work, we follow the dust model in H15, considering both the extinction from diffuse ISM \citep{1999A&A...350..381D} and from star-forming molecular clouds \citep{2000ApJ...539..718C}. The optical depth of dust, as a function of wavelength, is independently computed for each component; and a slab geometry is assumed to compute the total extinction of the relevant populations. 

Firstly, we calculate the extinction caused by the diffuse ISM. The wavelength dependent optical depth of galaxy disks is assumed as follows:
\begin{equation}
\label{eq:tauism}
\begin{aligned}
\tau_\lambda^{I S M}= & \left(\frac{A_\lambda}{A_v}\right)_{Z_{\odot}}(1+z)^{-1}\left(\frac{Z_{\mathrm{gas}}}{Z_{\odot}}\right)^s \\
& \times\left(\frac{\left\langle N_H\right\rangle}{2.1 \times 10^{21} \text { atoms cm }^{-2}}\right),
\end{aligned}
\end{equation}
where $\left(\frac{A_\lambda}{A_v}\right)_{Z_{\odot}}$ is the extinction curve for solar metallicity taken from \citet{1983A&A...128..212M}, $(1+z)^{-1}$ represents the redshift dependence, $Z_{\mathrm{gas}}$ is the metallicity of cold gas, $s = 1.35$ for $\lambda < 2000 \AA$ and $s = 1.6$ for $\lambda > 2000 \AA$. 
$\left\langle N_H\right\rangle$ is the mean column density of hydrogen:
\begin{equation}
    \left\langle N_H\right\rangle=\frac{M_{\text {cold }}}{1.4 m_p \pi\left(a R_{\text {gas }, \mathrm{d}}\right)^2} \text { atoms cm }^{-2},
\end{equation}
where $R_{\text {gas }, \mathrm{d}}$ is the scale-length of the cold gas disk, and $a = 3.36$. It should be noted that in previous work \citep{2011MNRAS.413..101G, 2015MNRAS.451.2663H}, although they claimed to adopt $a = 1.68$, the actual factor used in their code was approximately $a \sim 3.36$. Therefore, we use 3.36 instead of 1.68 in our calculations. 

\begin{figure*}
\centering
\includegraphics[width=2\columnwidth]{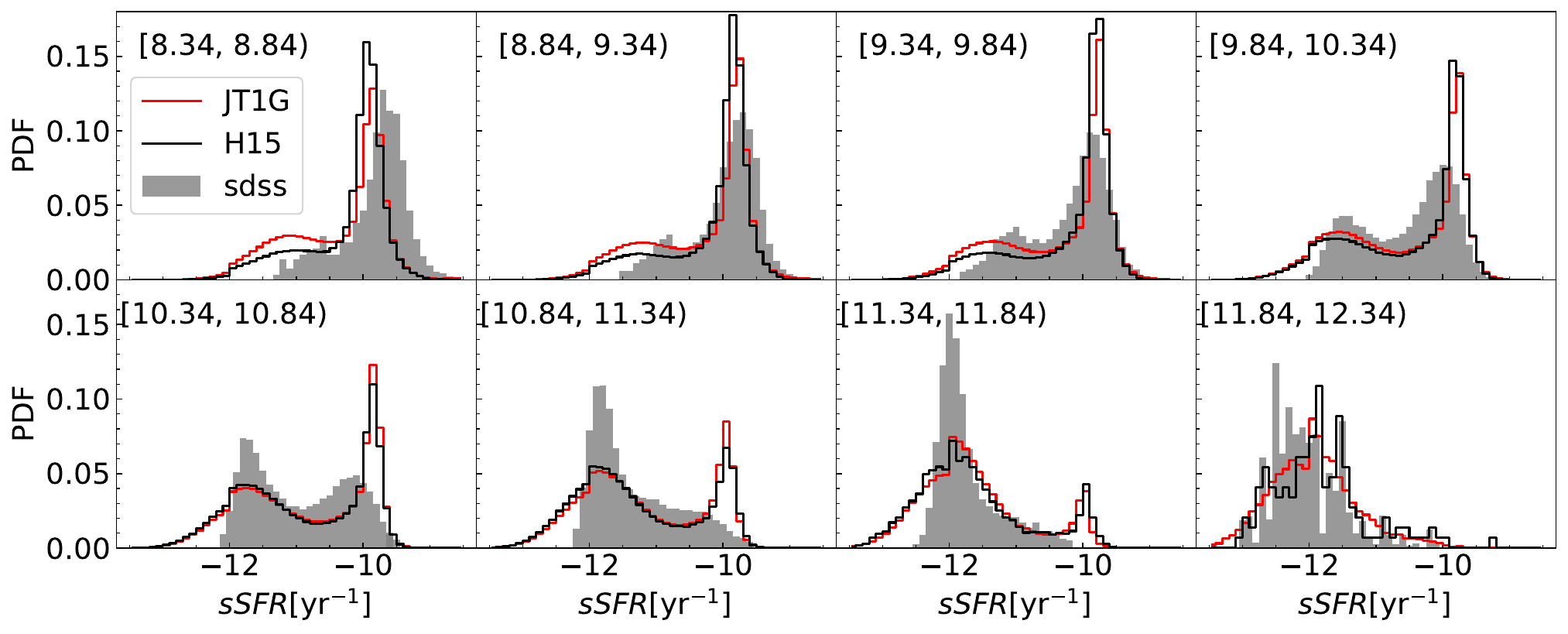}
\caption{Probability Distribution Function (PDF) of specific star formation rate for different stellar mass bins at redshift 0. The stellar mass ranges are shown in the upper-left corner of each panel. Gray shaded regions represent SDSS DR7 results \citep{2004MNRAS.351.1151B,2007ApJS..173..267S}, black lines depict H15 results and red lines are the results from our new model.}
\label{fig:ssfrpdf}
\end{figure*}

\begin{figure}
\centering
\includegraphics[width=0.8\columnwidth]{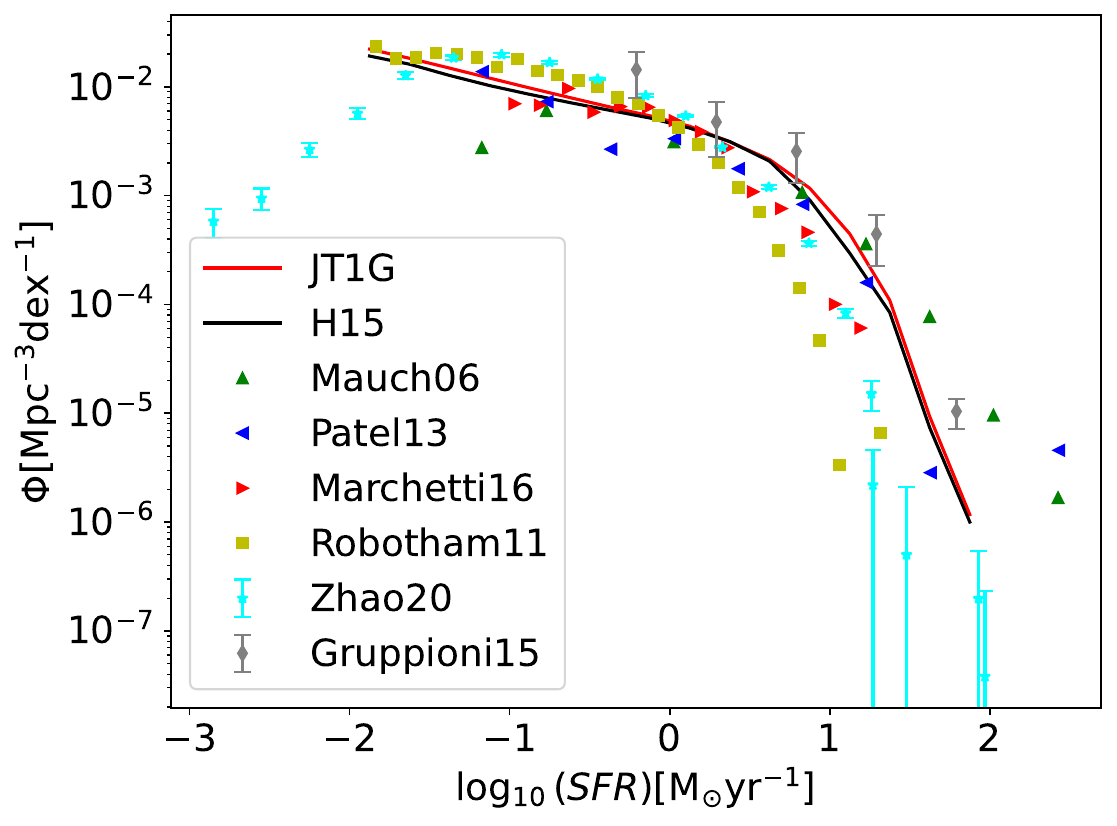}
\caption{Star Formation Rate Function at redshift 0. Various observational data points are included \citep{2007MNRAS.375..931M,2011MNRAS.413.2570R,2013MNRAS.428..291P,2015MNRAS.451.3419G,2016MNRAS.456.1999M,2020RAA....20..195Z}. }
\label{fig:sfrf}
\end{figure}

Another source contributing to the extinction is the molecular cloud around young stars. Following \citet{2000ApJ...539..718C}, we assume that only young stars born within the last 10Myr will suffer from such effects. The optical depth is calculated as follows:
\begin{equation}
    \tau_\lambda^{B C}=\tau_\lambda^{\mathrm{ISM}}\left(\frac{1}{\mu}-1\right)\left(\frac{\lambda}{5500 \AA}\right)^{-0.7},
\end{equation}
where $\mu$ is given by a Gaussian distribution with a mean of 0.3 and a standard deviation of 0.2, truncated at the boundaries of 0.1 and 1. 

The final extinction in magnitude for each component is given by:
\begin{equation}
    A_\lambda=-2.5 \log \left(\frac{1-\exp ^{-\tau_\lambda \sec \theta}}{\tau_\lambda \sec \theta}\right),
\end{equation}
where $\theta$ is the inclination angle between the angular momentum of the disk and the z-direction of the simulation box, and $\tau_\lambda$ is the optical depth of the corresponding component. Young stars (age less than 10Myr) suffer from both extinction components, while older stars are affected by diffuse ISM only. In the case of emission lines, we only consider the extinction from molecular clouds, as we only calculate the emission of star-forming regions.

In the following sections, all results are calculated using the new model and parameter settings applied to the merger trees from JT1G, unless otherwise specified.

\section{General galaxy properties}
\label{sec:galpred}

This section presents a comprehensive comparison of various galaxy properties between our model predictions and observational data. We categorise the comparison into two classes: one for properties utilised as constraints in our MCMC parameter adjustment, and properties directly related; and the other for properties that are not utilised as constraints.

\begin{figure*}
\centering
\includegraphics[width=1.5\columnwidth]{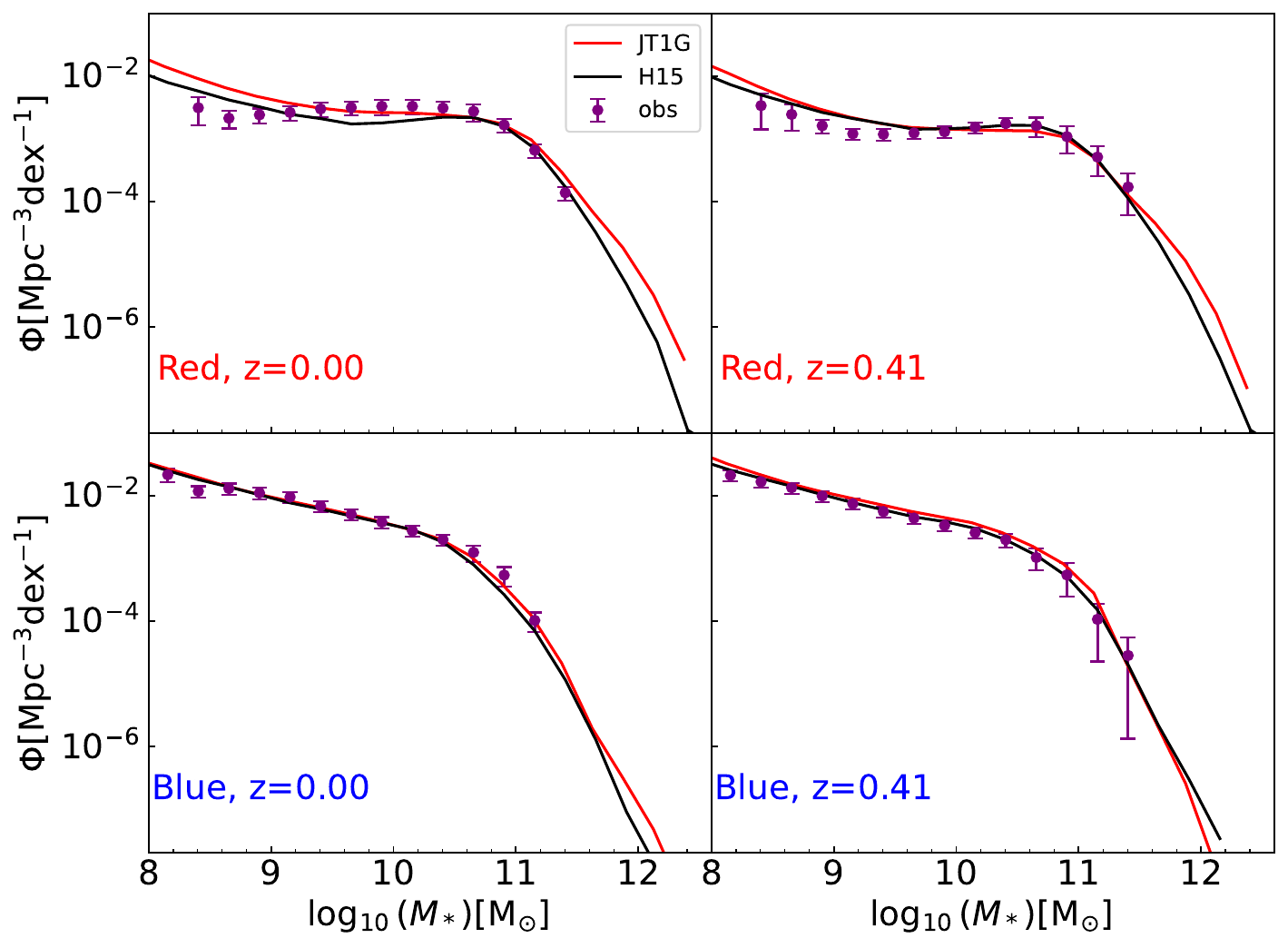}
\caption{Upper panels: SMF of red galaxies. Lower panels: SMF of blue galaxies. The left column corresponds to redshift 0, and the right column to redshift 0.4. Observational data points are sourced from \citet{2003ApJS..149..289B,2004ApJ...600..681B,2013ApJ...777...18M}, and \citet{2013AA...556A..55I}.}
\label{fig:rbmf}
\end{figure*}

\subsection{Comparison with observational galaxy properties used as MCMC constraints}
\subsubsection{Galaxy stellar mass functions}

Fig.~\ref{fig:smf} shows the galaxy stellar mass functions span redshifts from 0 to 3. The red lines depict our model results, while the black lines represent the results of MS using default H15 model. The purple symbols with error bars are the observational data points used in our MCMC procedures that were produced by combining various observational studies in an attempt to estimate systematic uncertainties in the constraints. Further details can be found in Appendix A of H15. Our predicted SMF aligns successfully with observational data across the entire stellar mass range, spanning from local to high redshift. Notably, at very low masses, our results exhibit a slightly steeper slope compared to observations. Both the new set of parameters and the higher resolution of the JT1G dark matter simulation compared to MS, where H15 was conducted, could potentially contribute to the observed differences. Our results outperform H15 for stellar masses beyond the knee of the SMF. This is because we have a much larger volume, eight times larger to be exact, which enables us to include more galaxies with high masses.

\subsubsection{Star formation rate}
The star formation rate (SFR) represents a fundamental statistical measure within galaxy formation theory. In this work, a passive galaxy is defined as having an sSFR less than $10^{-11}\rm yr^{-1}$, where ${\rm sSFR} = {\rm SFR}/M_*$ denotes the specific star formation rate. We utilise the passive fraction at z = 0.4 as a constraint when adjusting the parameters. 

Fig.~\ref{fig:ssfrpdf} displays the probability distribution function (PDF) of sSFR for different stellar mass bins at redshift 0. In accordance with H15, we assign a random Gaussian sSFR centred at ${\rm log}({\rm sSFR}) = -0.3\log(M_*) - 8.6$ with a dispersion of 0.5 for model galaxies with ${\rm log}({\rm sSFR}) \leq -12 \rm yr^{-1}$. The shaded regions represent results from SDSS DR7 \citep{2004MNRAS.351.1151B,2007ApJS..173..267S}, while the black lines are results from H15. Our new model predicts an sSFR distribution similar to H15 across the entire stellar mass range and shows a higher peak in sSFR for the subset of star-forming galaxies at low masses, which is more consistent with the SDSS data compared to H15.

In Fig.~\ref{fig:sfrf}, we show the SFRF at redshift 0, with various observational data points \citep{2007MNRAS.375..931M,2011MNRAS.413.2570R,2013MNRAS.428..291P,2015MNRAS.451.3419G,2016MNRAS.456.1999M,2020RAA....20..195Z}. It is noteworthy that there exists a significant discrepancy between different observational datasets, with variations of up to three orders of magnitude at the high SFR end. Our model results fall within the range covered by these observations, emphasising the need for more accurate measurements to better constrain theoretical models.

\subsubsection{Red and blue galaxies}

sSFR is closely correlated with galaxy colours, which are widely employed to distinguish the star formation states of galaxies. Therefore, we further study the stellar mass functions of red galaxies and blue galaxies at z = 0 and 0.4. We use the same colour cut as H15 for segregating galaxies into red and blue by: $u-r=1.85-0.075\times {\rm tanh}(({\rm M_{r}} + 18.07)/1.09)$. The upper and bottom panels in Fig.~\ref{fig:rbmf} show the SMF of red and blue galaxies, and the left and right columns present the findings at redshifts 0 and 0.4, respectively. Both our model results and those of H15 successfully reproduce the overall shape of the observations \citep{2003ApJS..149..289B,2004ApJ...600..681B,2013ApJ...777...18M,2013AA...556A..55I}. However, our model has slightly more red galaxies at the low mass end at z=0.  Furthermore, the SMF of red galaxies experiences a noticeable decline at $M_*\sim 10^{10}\Msun$ for H15 at z = 0, which is absent in our model. 

\begin{figure}
\centering
\includegraphics[width=0.8\columnwidth]{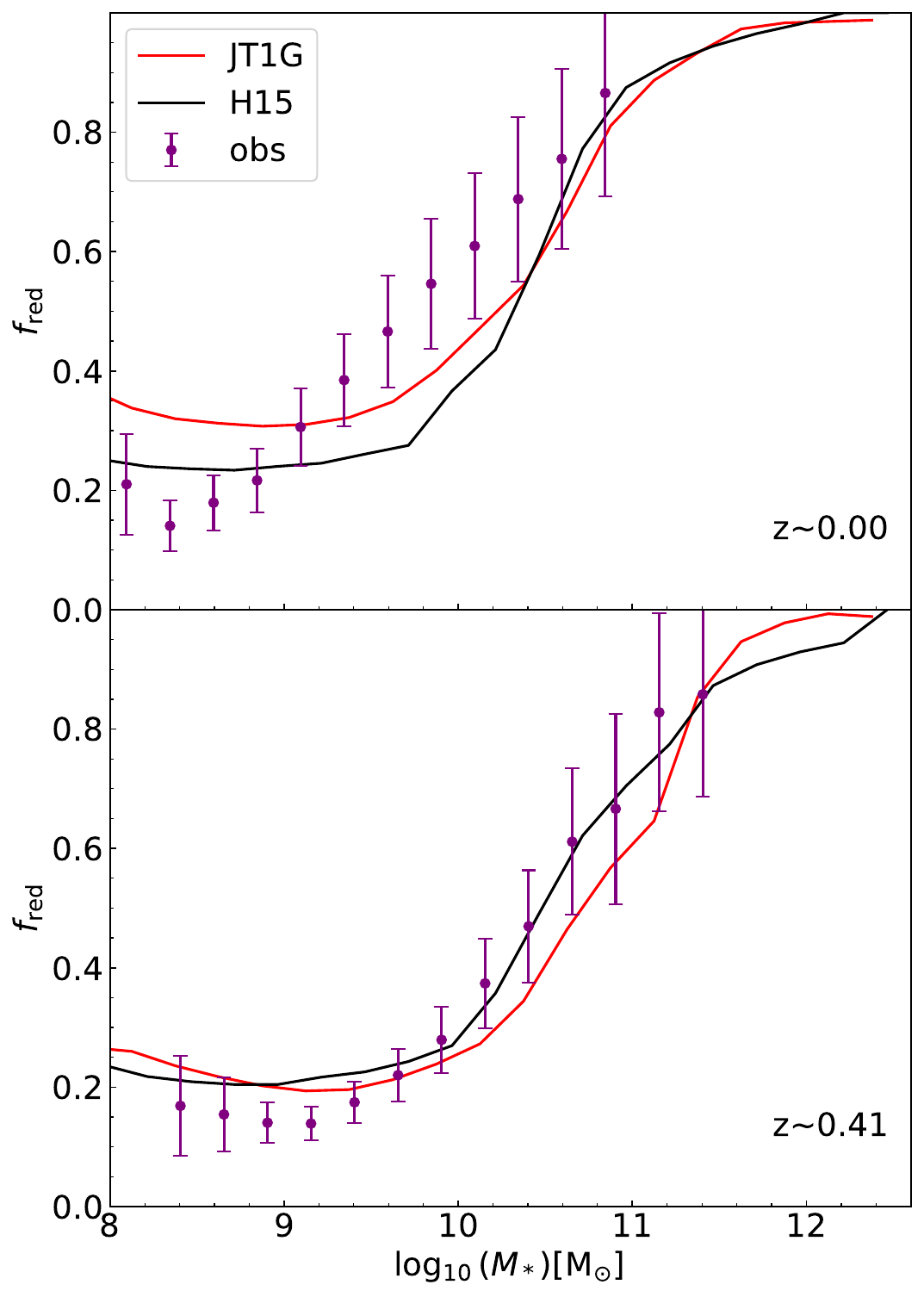}
\caption{Red fraction as a function of stellar mass. The upper panel corresponds to redshift 0, and the bottom panel to redshift 0.4. Observational data points are sourced from \citet{2003ApJS..149..289B,2004ApJ...600..681B,2013ApJ...777...18M}, and \citet{2013AA...556A..55I}.}
\label{fig:fred}
\end{figure}

The discrepancy between model predictions and observations becomes more pronounced when expressed in terms of the red fraction as a function of stellar mass (see Fig.~\ref{fig:fred}). The upper and bottom panels illustrate the red fraction as a function of stellar mass at redshift 0 and 0.4, respectively, with observational data points sourced from \citet{2003ApJS..149..289B,2004ApJ...600..681B,2013ApJ...777...18M} and \citet{2013AA...556A..55I}. At redshift 0, our model's results align marginally with the observations. We overestimate the proportion of red galaxies at the lower mass end and underestimate the red fraction at intermediate masses. H15 performs better at low masses but deviates more from the observations at $M_*\sim 10^{10}\Msun$. At redshift 0.4, both our model and H15 are in line with the observations, although at intermediate masses, our model exhibits slightly fewer passive galaxies compared to H15. These findings underscore the need for further investigation into the quenching mechanisms, particularly around the knee of the stellar mass function.

\subsubsection{Supermassive black holes}

\begin{figure}
\centering
\includegraphics[width=0.8\columnwidth]{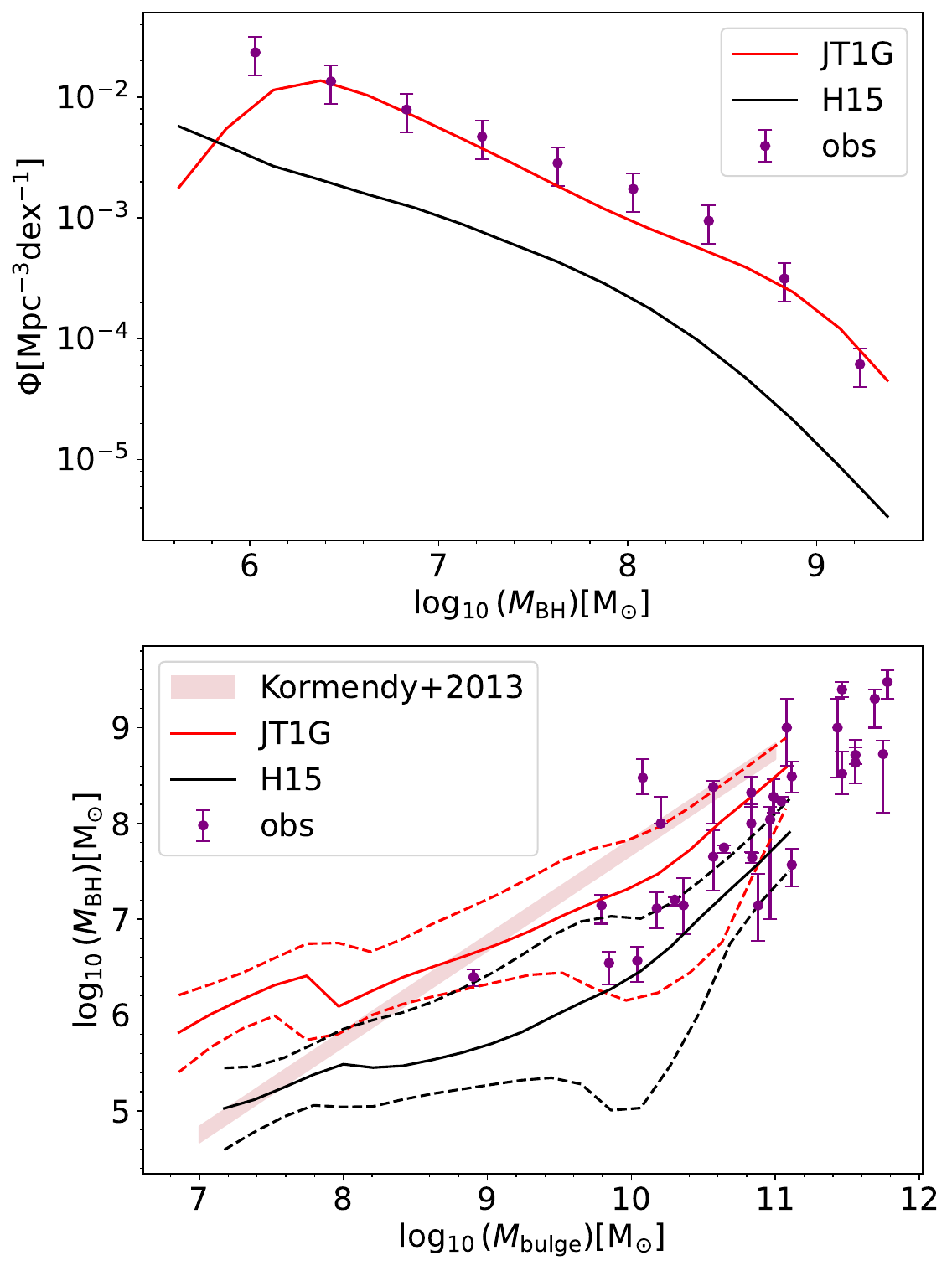}
\caption{Upper panel: Black Hole Mass Function at redshift 0. Purple dots represent the observed results of \citet{2009ApJ...690...20S}. Lower panel: Black Hole mass versus bulge mass relation. Observational data points from \citet{2004ApJ...604L..89H} (purple dots) and the relation from \citet{2013ARA&A..51..511K} (pink shaded region) are shown for comparison. The solid and dashed lines represent the median and 16\%(84\%) values.}
\label{fig:bh}
\end{figure}

It has been noted that H15 underestimates the mass of SMBH by approximately an order of magnitude \citep{2018ApJ...864...51W, 2019MNRAS.483..503Y}, suggesting that in the default H15 model the black hole growth rate is too slow. As illustrated in the upper panel of Fig.~\ref{fig:bh}, our model agrees well with the observed BHMF taken from \citet{2009ApJ...690...20S}, which is used in the MCMC procedures, while H15 significantly underestimates the BHMF by an order. The new parameter set significantly enhances the growth rate during mergers, while also reducing the efficiency of AGN feedback. This adjustment allows the SMBH to attain larger masses, suppressing star formation activities in massive galaxies. Modifications in these aspects were essential to achieve a more precise representation of the observed BHMF.

The bottom panel of Fig.~\ref{fig:bh} illustrates the SMBH mass versus bulge mass relation. Observational data is obtained from \citet{2004ApJ...604L..89H}, and the pink shaded region represents the relation from \citet{2013ARA&A..51..511K}. Our new model predicts larger SMBH masses at a given bulge mass than H15 and aligns more closely with the observations. This alignment is a direct consequence of our enhanced growth rate during mergers.

Some earlier studies also focus on improving the modeling of SMBH growth in \lgalaxies{}. \citet{2020MNRAS.495.4681I} introduces a time delay for Eddington accretion, promotes galaxy mergers, and incorporates additional pathways for SMBH growth, like disc instabilities. \citet{2023MNRAS.518.4672S} improves the modelling of SMBH seeds through various formation channels. \citet{2024arXiv240110983I} constructs a new framework by including the super-Eddington accretion events. All these advancements have resulted in better alignment of SMBH in \lgalaxies{} with observational data.

\subsection{Comparison with observational galaxy properties not used as MCMC constraints}
\subsubsection{Galaxy vs. dark halo relations}
\begin{figure}
\centering
\includegraphics[width=0.8\columnwidth]{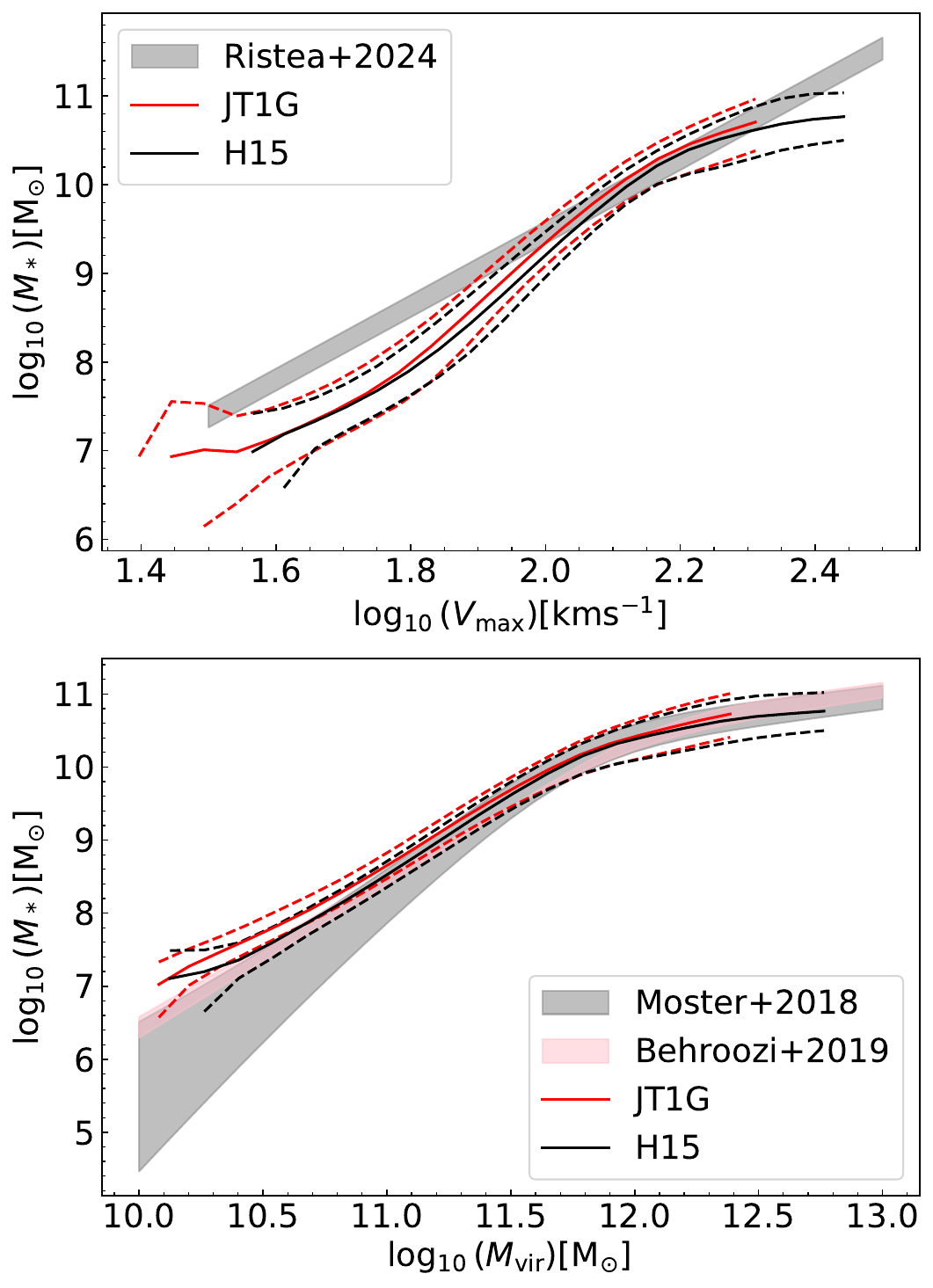}
\caption{Upper panel: the maximum velocity - stellar mass relation, compared with local results from SDSS-MaNGA \citep{2024MNRAS.527.7438R}. The solid and dashed lines represent the median and 16\%(84\%) values. Bottom panel: the virial mass - stellar mass relation, compared with results from abundance matching methods \citep{2018MNRAS.477.1822M,2019MNRAS.488.3143B}. The solid and dashed lines represent the median and 16\%(84\%) values.} The maximum velocity and virial masses were determined at $z = 0$ for central galaxies and at the last infall time for satellite galaxies.
\label{fig:gh}
\end{figure}

The relationship between galaxy and dark halo properties represents one of the most fundamental connections in the field of galaxy formation. Fig.~\ref{fig:gh} illustrates these scaling relations by highlighting the correlation between galaxy stellar mass and both the maximum halo velocity and the maximum virial mass. The maximum velocity and virial masses were determined at z = 0 for central galaxies, while for satellite galaxies, they were determined at the last infall time. In general, both of these scaling relations demonstrate good agreement with the H15 models, particularly in terms of their slopes. With regard to the correlation between stellar mass ($M_*$) and virial mass ($M_{\rm vir}$), the newly proposed model showcases somewhat enhanced stellar masses for JT1G galaxies in comparison to the H15 models. This suggests a higher galaxy formation efficiency in the new model. It is noteworthy to mention that despite the minor variations, both model forecasts agree with direct measurements obtained using local data \citep{2024MNRAS.527.7438R} and abundance matching methodologies \citep{2018MNRAS.477.1822M,2019MNRAS.488.3143B}.

\subsubsection{Gas metallicity}
\begin{figure}
\centering
\includegraphics[width=0.8\columnwidth]{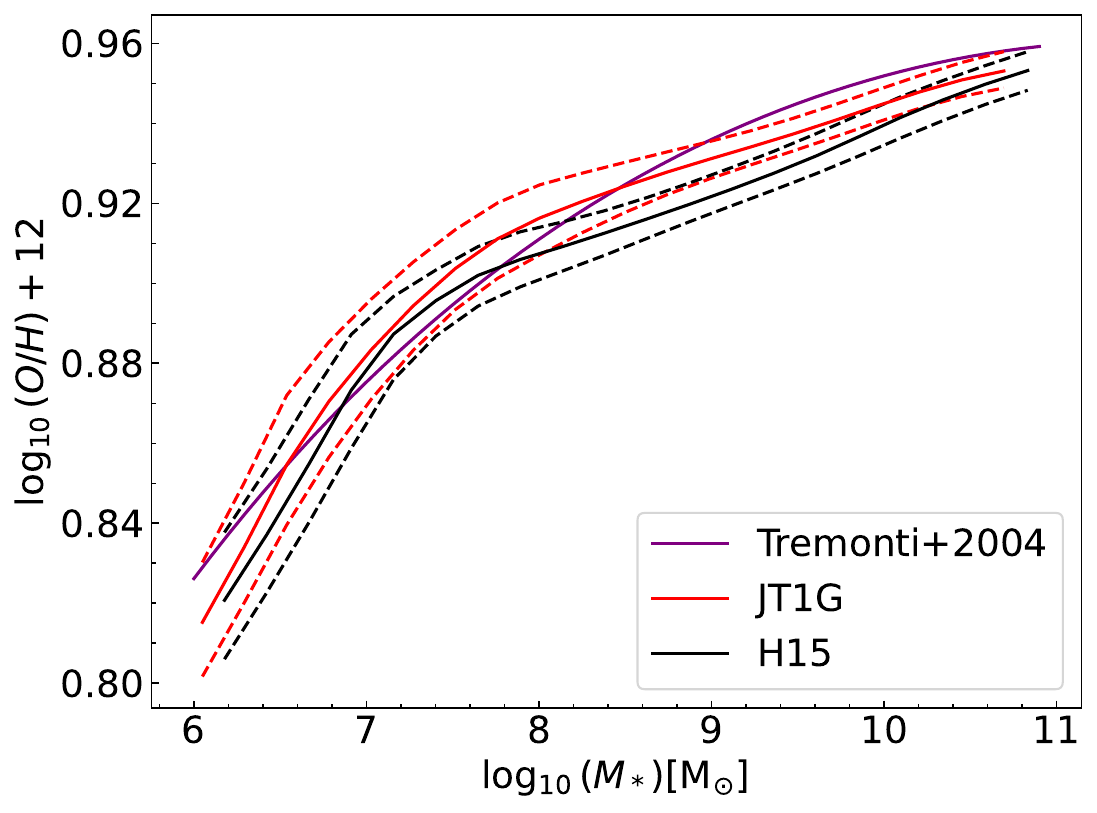}
\caption{The relationship between metallicity and stellar mass. The solid and dashed lines represent the median and 16\%(84\%) values. The gas metallicity ($Z_{\text{gas}}$) is determined by the ratio of the metal mass in cold gas to the mass of cold gas and converted to oxygen abundance using the formula $12 + \log_{10}({\rm O/H})_{\rm gas}= 8.69 + \log_{10}(Z_{\rm gas}/Z_{\odot})$. The purple line shows the result from \citet{2004ApJ...613..898T}.}
\label{fig:metal}
\end{figure}

Fig.~\ref{fig:metal} depicts the relationship between gas metallicity and stellar mass. The metallicity of the gas ($Z_{\rm gas}$) is determined by the ratio of the metal mass in the cold gas to the mass of the cold gas. Then it is converted to oxygen abundance using the formula $12 + \log_{10}({\rm O/H})_{\rm gas}= 8.69 + \log_{10}(Z_{\rm gas}/Z_{\odot})$. It is evident that our newly developed model exhibits somehow higher gas metallicity compared to H15 across a wide range of masses. This characteristic enables it to be more in line with the observations \citep{2004ApJ...613..898T} made at high masses. 

\subsubsection{Evolution of SFR}

\begin{figure}
\centering
\includegraphics[width=0.8\columnwidth]{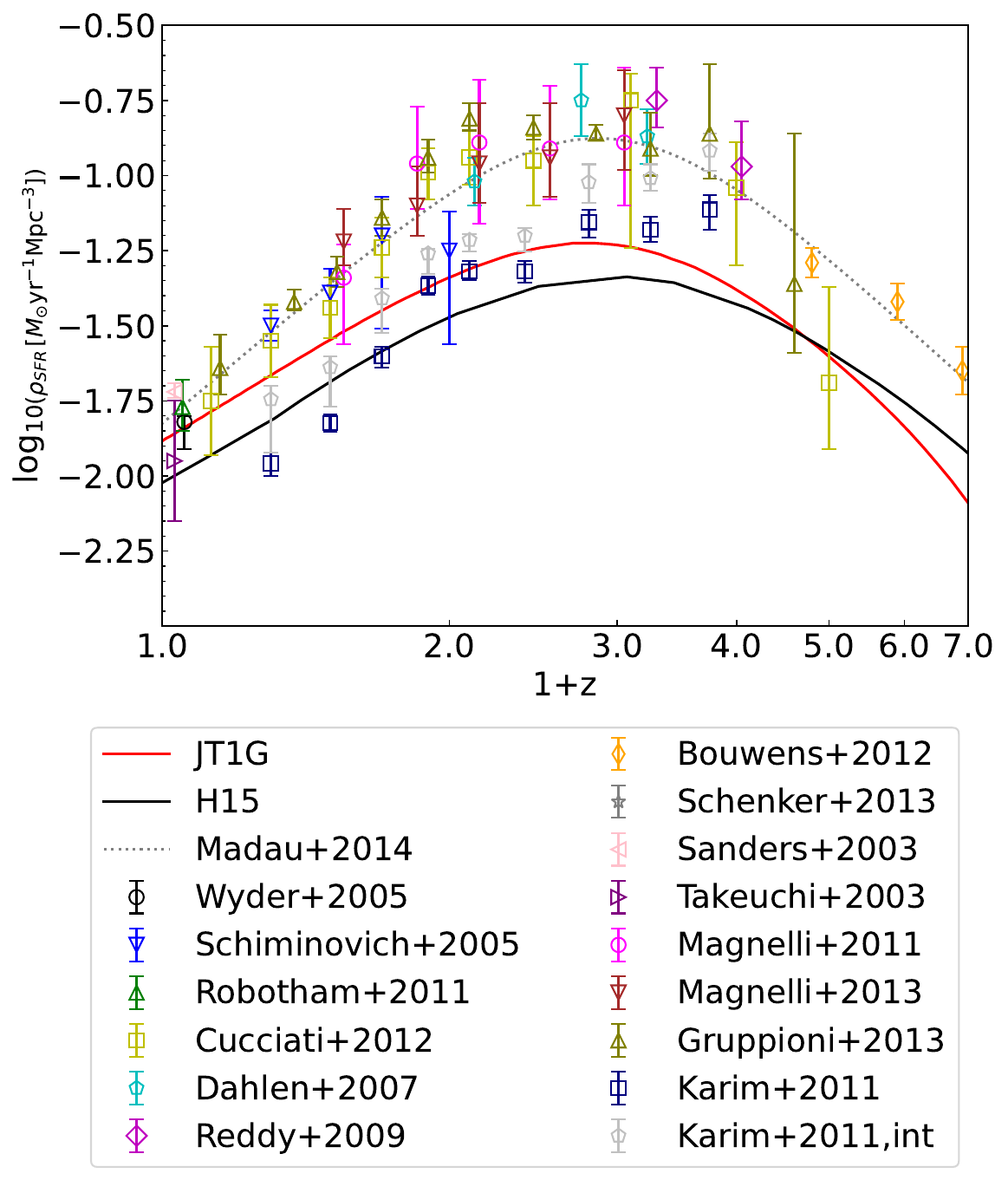}
\caption{Star Formation Rate Density as a function of redshift from 0 to 6. Various observational data points are included \citep{2003AJ....126.1607S,2003ApJ...587L..89T,2005ApJ...619L..47S,2005ApJ...619L..15W,2007ApJ...654..172D,2009ApJ...692..778R,2011ApJ...730...61K,2011MNRAS.413.2570R,2011A&A...528A..35M,2012ApJ...754...83B,2012A&A...539A..31C,2013MNRAS.432...23G,2013A&A...553A.132M,2013ApJ...768..196S}. The gray line represents the best fitting result from \citet{2014ARA&A..52..415M}.}
\label{fig:sfrd}
\end{figure}

In Fig.~\ref{fig:sfrd}, we compare the model-predicted star formation rate density (SFRD) as a function of redshift to various observational results \citep{2003AJ....126.1607S,2003ApJ...587L..89T,2005ApJ...619L..47S,2005ApJ...619L..15W,2007ApJ...654..172D,2009ApJ...692..778R,2011ApJ...730...61K,2011MNRAS.413.2570R,2011A&A...528A..35M,2012ApJ...754...83B,2012A&A...539A..31C,2013MNRAS.432...23G,2013A&A...553A.132M,2013ApJ...768..196S} from redshift 0 to 6. The gray line represents the best-fitting result from \citet{2014ARA&A..52..415M}. The observed SFRD peaks at a redshift of 2 gradually, which diminishes in magnitude when moving towards either higher or lower redshift values. The results of our simulation generally reproduce this trend, with specific values falling within the observational constraints. This suggests that our model effectively traces the evolutionary history of the SFR in the universe. More stars are formed in the new model on JT1G compared to H15 on MS below redshift 4.

\begin{figure*}
\centering
\includegraphics[width=2\columnwidth]{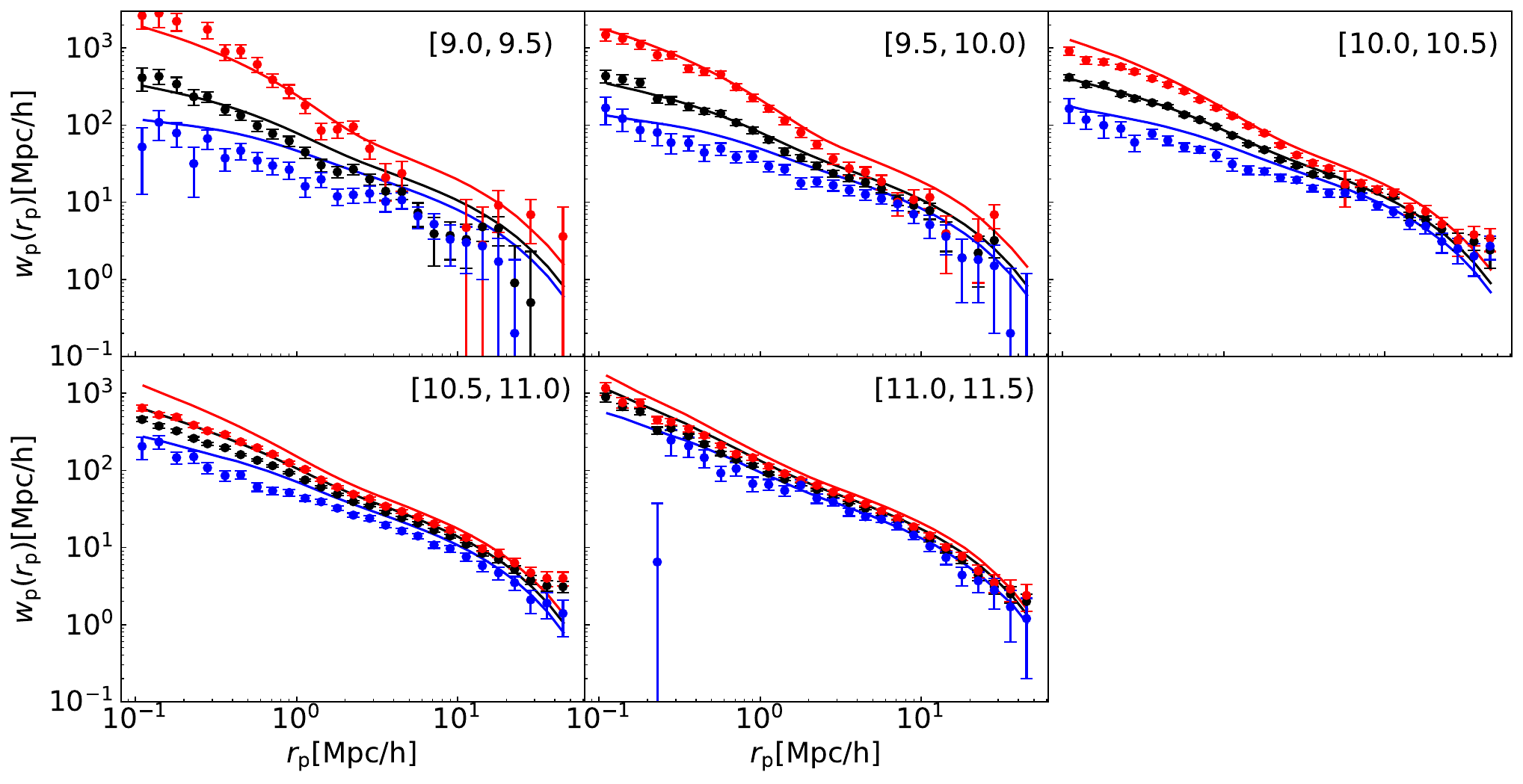}
\caption{Projected auto-correlation functions of simulated galaxies compared with observations in SDSS DR7. Different panels represent different stellar mass ranges, with black lines depicting results for the entire galaxy sample. The stellar mass ranges are shown in the
upper right corner of each panel. Red and blue lines represent results for red and blue galaxies, respectively. Symbols with error bars denote measurements from \citet{2006MNRAS.368...21L}.}
\label{fig:xi}
\end{figure*}

\begin{figure*}
\centering
\includegraphics[width=2\columnwidth]{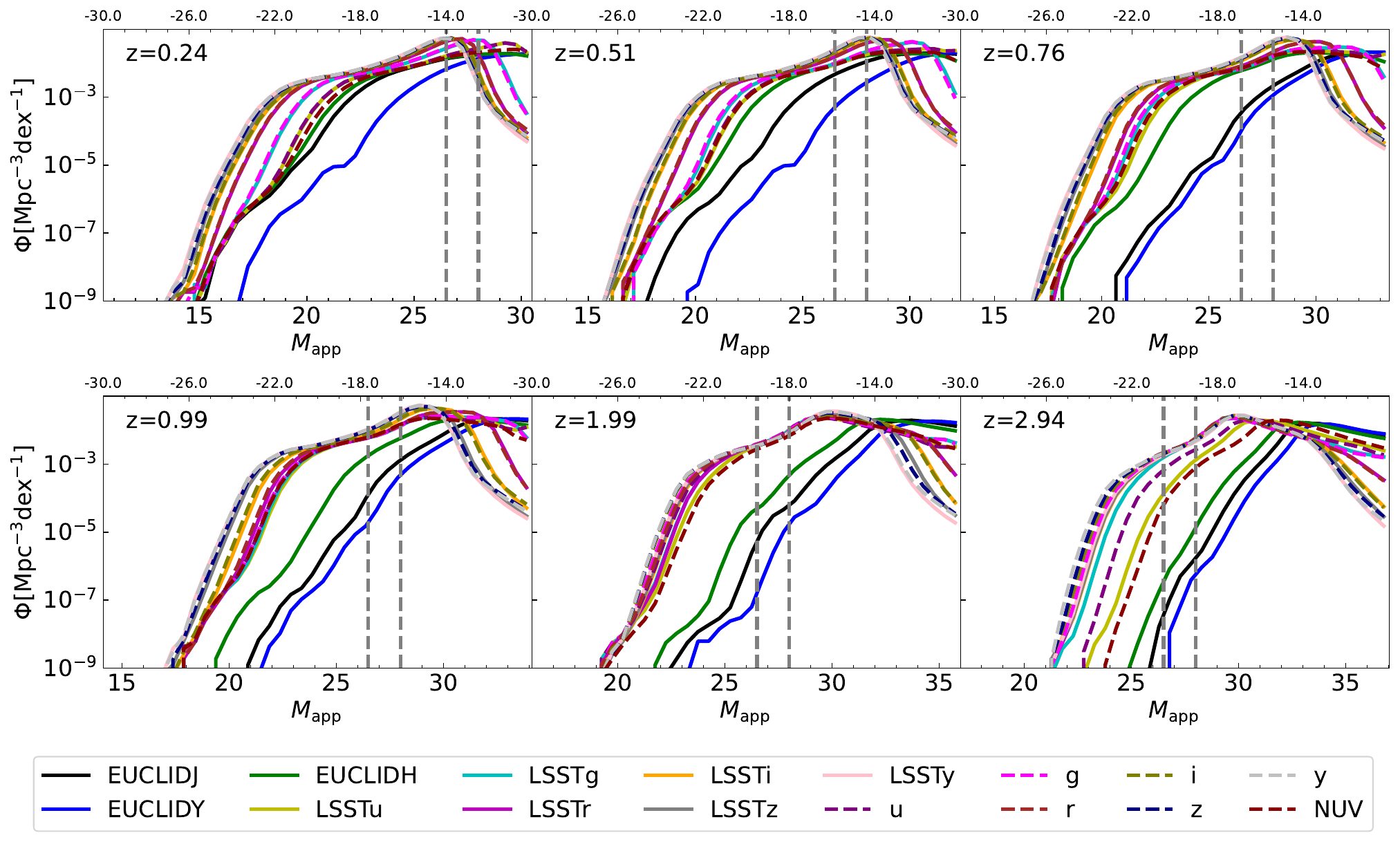}
\caption{Luminosity functions (LF) for various surveys in observed-frame across redshifts from 0 to 3. The panels show the LF for the $Y$, $J$, and $H$ bands of \textit{EUCLID}, and the $u$, $g$, $r$, $i$, $z$, and $y$ bands of LSST, and the $NUV$, $u$, $g$, $r$, $i$, $z$, and $y$ bands of CSST. All magnitudes account for dust extinction mentioned in Sec.~\ref{sec:dust}. The bottom x-axis displays apparent magnitude, while the luminosity distance is computed from the corresponding redshift. The top x-axis represents absolute magnitude. The two vertical dashed gray lines signify the detection limits of the CSST main survey (apparent magnitude of 26.5) and the deep field survey (apparent magnitude of 28). }
\label{fig:lf}
\end{figure*}

\begin{table}
	\centering
	\caption{Details of the $L_{\rm [OII]}$-selected sample and stellar mass-selected sample.}
	\label{tab:subsample}
	\begin{tabular}{lcccr} 
		\hline
             name & redshift & $\log_{10} M_*$ & $N_{\rm g}$ \\
		\hline
            M0 & 0.76 & [9.9, 10.2] & 4462522\\
            M1 & 0.76 & [10.2, 10.5] & 3197997\\
            M2 & 0.76 & [10.5, 10.9] & 2825084\\
            M3 & 0.76 & [10.9, $\inf$] & 894120\\
		\hline

		\hline
             name & redshift & $\log_{10} L_{\rm [OII]}$ & $N_{\rm g}$ \\
		\hline
            L0 & 0.76 & [40.85, 41.15] & 8925507\\
            L1 & 0.76 & [41.15, 41.45] & 7693206\\
            L2 & 0.76 & [41.45, 41.75] & 6101810\\
            L3 & 0.76 & [41.75, $\inf$] & 12783344\\
		\hline
	\end{tabular}
\end{table}

\subsubsection{Galaxy correlation functions}
The study of galaxy correlation functions encompasses the analysis of the spatial distribution of galaxies, which in turn provides valuable information on the distribution of matter. 
We use the Landy-Szalay estimator \citep{1993ApJ...412...64L,1998ApJ...494L..41S} to calculate the redshift-space cross-correlation functions between different subsamples:
\begin{equation}
    \xi_{x y}\left(r_{\mathrm{p}}, r_\pi\right)=\left[\frac{D_x D_y-D_x R_y-D_y R_x+R_x R_y}{R_x R_y}\right],
\end{equation}
where $r_{\mathrm{p}}$ is the distance along the projected direction and $r_\pi$ is distance along the line-of-sight. $D_x D_y, D_x R_y, D_y R_x,$ and $R_x R_y$ are the normalised pair counts for data x-data y, data x-random y, data y-random x and random x-random y. Here x, and y indicate different subsamples. When x=y it becomes an auto-correlation function. The real-space projected correlation function $w_{\mathrm{p}}^{x y}\left(r_{\mathrm{p}}\right)$ is calculated by integrating $\xi_{x y}\left(r_{\mathrm{p}}, r_\pi\right)$ along the line-of-sight \citep{1983ApJ...267..465D}:
\begin{equation}
    w_{\mathrm{p}}^{x y}\left(r_{\mathrm{p}}\right)=2 \int_0^{r_{\pi, \max }} \xi_{x y}\left(r_{\mathrm{p}}, r_\pi\right) d r_\pi,
\end{equation}
where $r_{\pi, \max } = 40\mpch$.
We adopt the same $r_{\rm p}$ and $r_\pi$ values as in \citet{2006MNRAS.368...21L} to facilitate a direct comparison of our results with theirs. Specifically, we employ 28 $r_{\rm p}$ bins spanning from 0.1-50 $\rm Mpc/h$ with equal logarithmic intervals and 40 $r_\pi$ bins covering from 0-40 $\rm Mpc/h$ with equal linear intervals. The Corrfunc code \citep{2020MNRAS.491.3022S} is utilised for the calculation of the correlation function.

Given that galaxy clustering can be strongly influenced by their mass, we have categorised our model galaxies into various bins based on their stellar masses. Fig.~\ref{fig:xi} illustrates the comparison between the projected autocorrelation functions of our simulated galaxies and those observed in SDSS DR7. Each panel within the figure represents a distinct stellar mass range, shown in the upper right corner, with the black lines representing the results for the entire galaxy sample. Notably, the results for red and blue galaxies are depicted by red and blue lines, respectively. The symbols accompanied by error bars in the figure signify the measurements obtained from \citet{2006MNRAS.368...21L}. 

Our catalogue effectively reproduces the projected autocorrelation function across the majority of stellar mass ranges while also capturing its dependence on colour. However, certain discrepancies between the simulation and the observations are evident. Specifically, at larger radii, we observe deviations for galaxies smaller than $10^{10}\Msun$. This could potentially be attributed to the limited volume occupied by these faint galaxies, thereby experiencing a notable cosmic variance effect. 

Furthermore, we note differences between the predictions of our model and the observations at smaller radii for galaxies exceeding $10^{10.5}\Msun$, indicating a significant proportion of satellite galaxies. By segregating red galaxies from blue galaxies, we find that the more strongly clustering of red galaxies contributes more prominently to the overall excess observed at smaller radii.

In summary, our refined model, incorporating adjusted parameters, excels in reproducing a wide range of observational properties: the stellar mass functions from local to redshift 3, the bimodal colour distributions of galaxies, the black hole mass functions, as well as other galaxy-halo relations and the clustering of galaxies, etc.

\section{Properties of Various Emission Lines}
\label{sec:emipred}

In this section, we first demonstrate that our galaxy catalogue meets the requirement for the next generation of large-scale surveys. We then present the model predictions of emission lines in comparison with observations. In Sec.~\ref{sec:emilf} we show the luminosity function of H${\alpha}$, [OII], [OIII], and [OIII] + H${\beta}$, comparing our model results with a set of observations from local to high redshift. In Sec.~\ref{sec:emixi}, we further compare the projected correlation function of $L_{\rm [OII]}$-selected samples with observations. Finally, in Sec.~\ref{sec:emibias}, we show the model predicted bias of galaxies as a function of luminosity for different emission lines.

\subsection{Luminosity functions for the next generation of large-scale surveys}
Predictions for the luminosity function (LF) of various surveys can provide more direct forecasts for future observations. We convolve the filter functions of various surveys with the simulated galaxy SED to obtain the observed-frame LF in different bands in Fig.~\ref{fig:lf}. It includes the luminosity functions in the \textit{Euclid} $Y$, $J$, and $H$ bands, in the LSST $u$, $g$, $r$, $i$, $z$, and $y$ bands, and in the CSST $NUV$, $u$, $g$, $r$, $i$, $z$, and $y$ bands across redshifts from 0 to 3. All magnitudes account for both the two components of dust extinction mentioned in Sec.~\ref{sec:dust} and include the contributions from various emission lines. The bottom x-axis displays apparent magnitude, while the top x-axis represents absolute magnitude. The two vertical dashed gray lines on the figures signify the detection limits of the CSST main survey (apparent magnitude of 26.5) and the deep field survey (apparent magnitude of 28). The transition point in the luminosity function, where the number density starts to decrease towards the dimmer end, serves as an approximate indicator of completeness. It happens at about 26.5 magnitudes at z = 0.24, and around 28 magnitudes at z = 0.5. Our simulation is deemed complete for the main survey above a redshift of 0.3 and the deep field survey above a redshift of 0.5. Similar completeness is expected for other surveys. We note that most tracers are above $z\sim$ 0.5. Meanwhile, the large box size of JT1G (1 Gpc/h) makes it suitable for studying large-scale structures. Therefore, our results offer a reasonably comprehensive prediction for the outcomes of the next generation of large-scale surveys.

\begin{figure*}
\centering
\includegraphics[width=2\columnwidth]{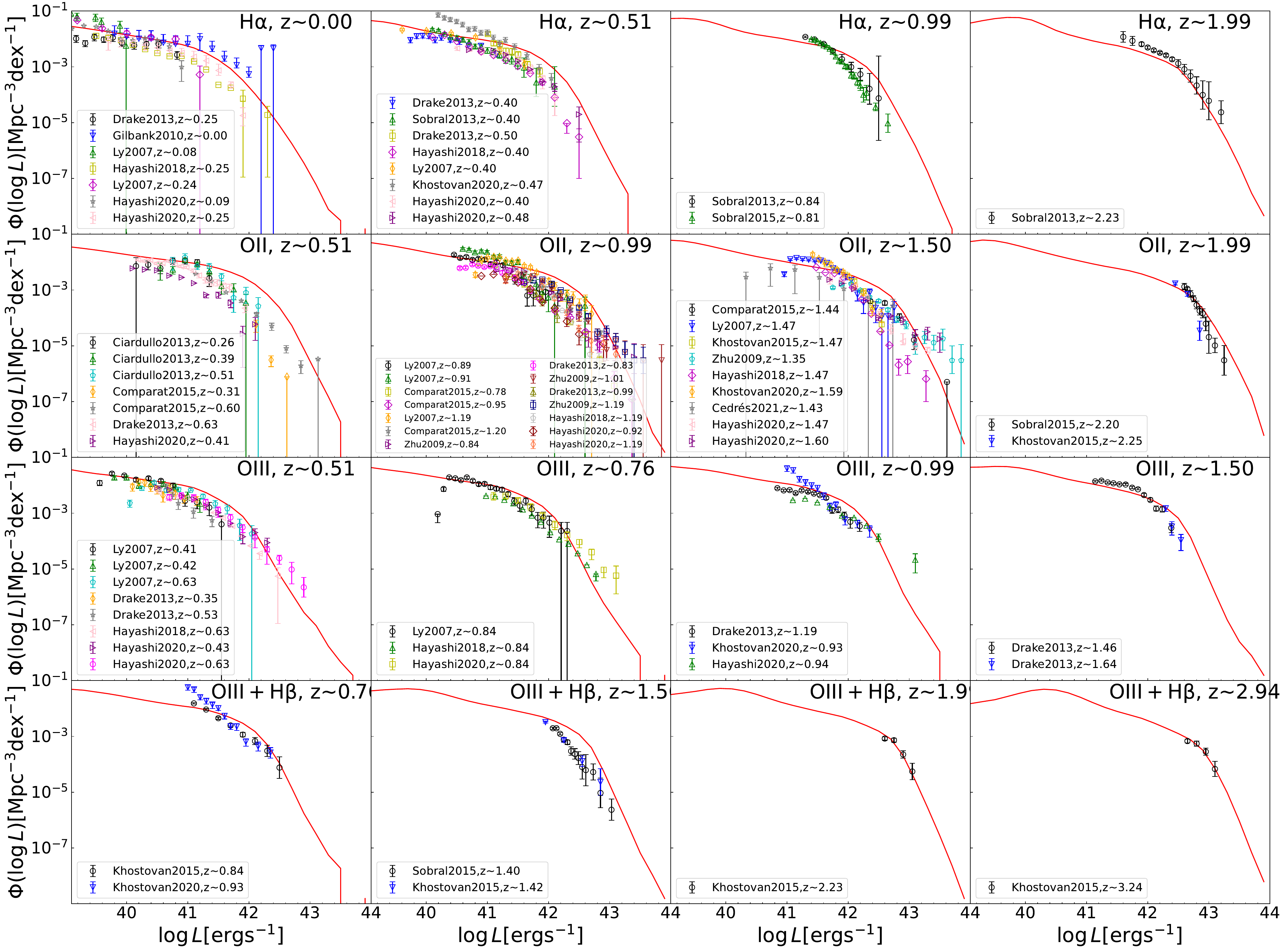}
\caption{Luminosity function of H$_\alpha$, [OII], [OIII], and [OIII] + H$\beta$ from redshift 0 to 3. The model predictions include dust attenuation from young birth cloud to match direct observational symbols. Each row corresponds to a different emission line, while each panel in the row represents a specific redshift. The first row displays the H$\alpha$ luminosity function at redshifts of approximately 0, 0.5, 1, and 2. The second row presents the [OII] doublet luminosity function at redshifts of approximately 0.5, 1, 1.5, and 2. The third row exhibits the [OIII] doublet luminosity function at redshifts around 0.5, 0.76, 1, and 1.5. The last row illustrates the luminosity function of [OIII]5007 + H$\beta$ at redshifts of approximately 0.75, 1.5, 2, and 3. Observational data points are taken from various studies \citep{2007ApJ...657..738L,2010MNRAS.405.2594G,2013MNRAS.433..796D,2013MNRAS.428.1128S,2015MNRAS.451.2303S,2009ApJ...701...86Z,2013ApJ...769...83C,2015A&A...575A..40C,2015MNRAS.452.3948K,2018PASJ...70S..17H,2020MNRAS.493.3966K,2020PASJ...72...86H,2021A&A...649A..73C}. Despite the general good agreement, there are slight discrepancies, particularly at the low-redshift knee of the [OII] luminosity function.}
\label{fig:emilf}
\end{figure*}

\begin{figure*}
\centering
\includegraphics[width=2\columnwidth]{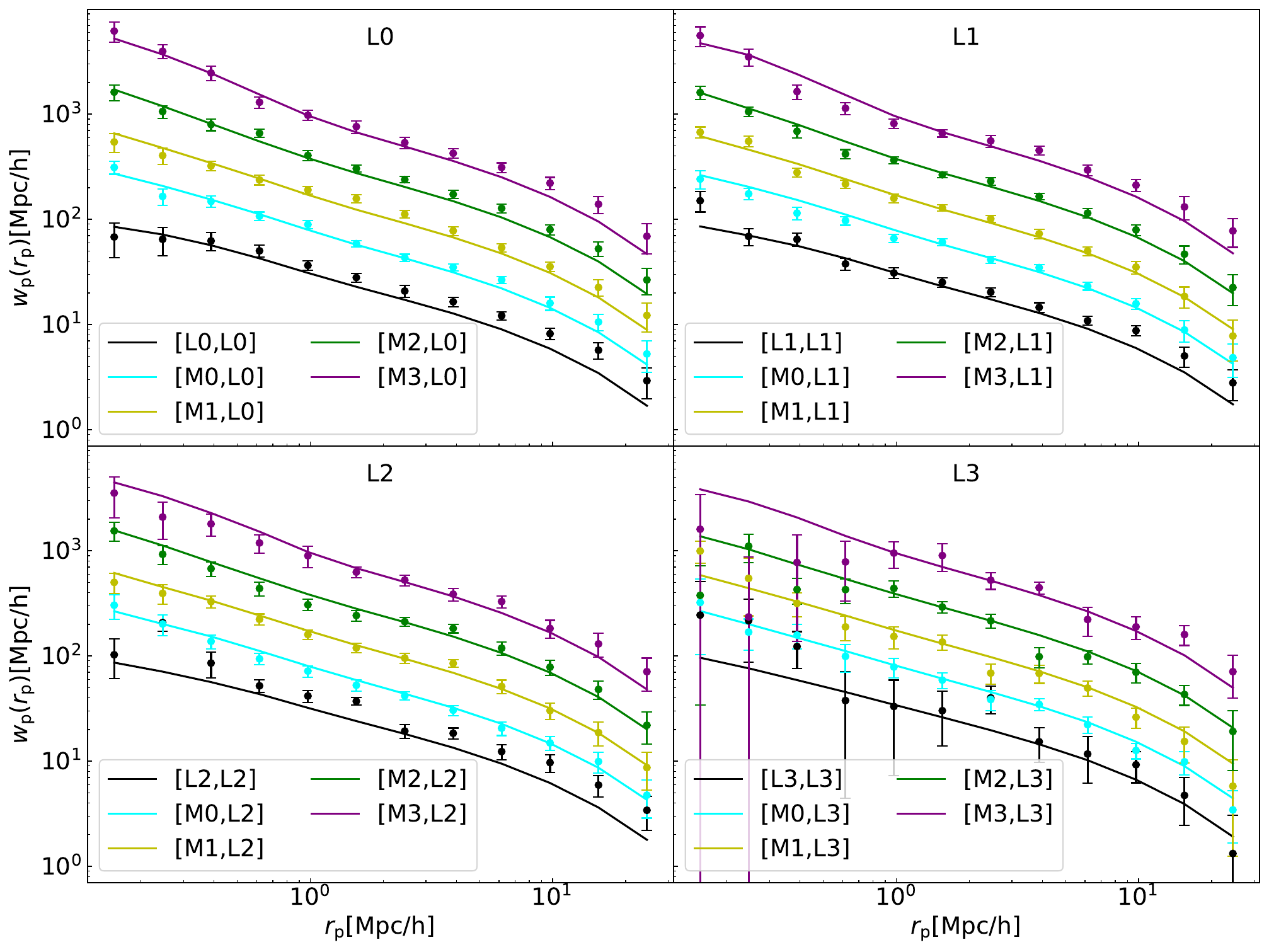}
\caption{Real-space cross (auto) correlation functions for subsamples L0, L1, L2, and L3 at redshift 0.76 presented in different panels. Observed data points with error bars are from \citet{2022ApJ...928...10G}, derived from VIPERS at redshift 0.5 - 0.8. In each panel, the black line represents the autocorrelation function of the specified [OII] luminosity subsample. Cyan/yellow/green/purple lines show the cross-correlation functions between the given [OII] luminosity-selected subsamples and stellar mass-selected subsamples M0/M1/M2/M3. These cross-correlation lines are shifted by a factor of $2^{\rm n}$, where n varies with colour (n=1(cyan), n=2(yellow), n=3(green), n=4(purple)) for clear illustration.}
\label{fig:emixi}
\end{figure*}

\subsection{Luminosity Function of Various Emission Lines}
\label{sec:emilf}

Fig.~\ref{fig:emilf} shows the luminosity function of H$\alpha$, [OII], [OIII], and [OIII]+H$\beta$ from redshift 0 to 3. The model predictions incorporate dust attenuation from young birth cloud as mentioned in Sec.~\ref{sec:dust} to align with direct observational symbols obtained from various sources. For clarity, we selected four representative redshifts to display for each line, ensuring the inclusion of corresponding observational data. Each observational point is presented in panels corresponding to the nearest redshift values. 

The first row in Fig.~\ref{fig:emilf} presents the LF of H$\alpha$ at z $\sim$ 0, 0.5, 1, and 2. Observational data points are collected from \citet{2007ApJ...657..738L,2010MNRAS.405.2594G,2013MNRAS.433..796D,2013MNRAS.428.1128S,2015MNRAS.451.2303S,2018PASJ...70S..17H,2020MNRAS.493.3966K,2020PASJ...72...86H}. Our model effectively reproduces the observed H$\alpha$ LF across a wide luminosity spectrum, ranging from $10^{39} \ergs$ to $10^{43} \ergs$, covering redshifts from 0 to 2. Since the luminosity of H$_\alpha$ is directly related to the SFR by Eq.~\ref{eq:ha_sfr}, the successful reproduction of the H$_\alpha$ LF suggests that our model can accurately emulate the overall SFRF from redshift 0 to 2.

The second row in Fig.~\ref{fig:emilf} displays the LF of [OII] at z $\sim$ 0.5, 1, 1.5, and 2. The [OII] luminosity is computed as the sum of the [OII] doublet. Observed data points are collected from \citet{2007ApJ...657..738L, 2009ApJ...701...86Z, 2010MNRAS.405.2594G, 2013ApJ...769...83C, 2013MNRAS.433..796D, 2015MNRAS.451.2303S, 2015A&A...575A..40C, 2015MNRAS.452.3948K,2018PASJ...70S..17H,2020MNRAS.493.3966K,2020PASJ...72...86H,2021A&A...649A..73C}. The overall agreement between observations and simulation is satisfactory, although there is a slight overestimation of the LF at the knee at low redshift. It is worth noting that limitations arise from the survey volume, which only ranges from tens of thousands to several hundred thousand $\rm Mpc^3$. Due to this constraint, the survey is susceptible to cosmic variance at these luminosities.

The third row in Fig.~\ref{fig:emilf} illustrates the LF of the [OIII] doublet at z $\sim$ 0.5, 0.76, 1, and 1.5. Observational data points are collected from \citet{2007ApJ...657..738L,2013MNRAS.433..796D,2018PASJ...70S..17H,2020MNRAS.493.3966K,2020PASJ...72...86H}. The last row shows the luminosity function of [OIII]5007+H$\beta$ at z $\sim$ 0.75, 1.5, 2, and 3, with observational points collected from \citet{2015MNRAS.451.2303S, 2015MNRAS.452.3948K,2020MNRAS.493.3966K}. Our model predictions demonstrate consistency with the observed [OIII] doublet and [OIII]5007+H$\beta$ LFs within a wide range of luminosity and redshift.

In summary, our study successfully reproduces the observed luminosity function for a set of emission lines from the local universe to redshift 3. This achievement suggests that both our semi-analytic galaxy model and the emission line model align with the actual universe.

\subsection{Clustering of ELG}
\label{sec:emixi}

We investigate the clustering of emission line galaxies by calculating their projected auto- and cross-correlation functions for various subsamples. Following the methodology outlined in \citet{2022ApJ...928...10G}, we partition our simulated galaxies at redshift 0.76 into distinct subsamples based on their [OII] luminosity and stellar mass. In practice, we create four samples based on [OII] luminosity, denoted as $L0$, $L1$, $L2$, and $L3$. 
For the cross-correlation functions, we generate four samples based on their stellar masses, denoted as $M0$, $M1$, $M2$, and $M3$. Detailed information is provided in Table ~\ref{tab:subsample}. 

We present the projected auto- and cross-correlation functions in Fig.~\ref{fig:emixi}. Each panel is for one particular [OII] luminosity bin. Observed data points with error bars are obtained from \cite{2022ApJ...928...10G}, derived from the VIMOS Public Extragalactic Redshift Survey \citep[VIPERS,][]{2018A&A...609A..84S}. In each panel, the black line represents the auto-correlation function of the specified [OII] luminosity sample. Our model successfully replicates the auto-correlation functions of galaxies within different [OII] luminosity bins across a broad range of radii, including one-halo and two-halo terms. The cross-correlation functions between the subsamples selected based on [OII] luminosity and subsamples selected based on stellar mass are depicted by the cyan/yellow/green/purple lines. It should be noted that these cross-correlation lines have been appropriately shifted by a factor of $2^n$, where n corresponds to the specific designation of the sample, M0, M1, M2, M3. Results from different stellar mass selected samples are presented using different colours as denoted in the bottom left corner of each panel. Overall, our model predictions exhibit an excellent agreement with observations for all subsamples. The only exception appears for the projected cross-correlation functions between M3 and L3, where the model prediction is higher at small scales compared to observations, implying a stronger star formation close to massive central galaxies.

\begin{figure*}
\centering
\includegraphics[width=2\columnwidth]{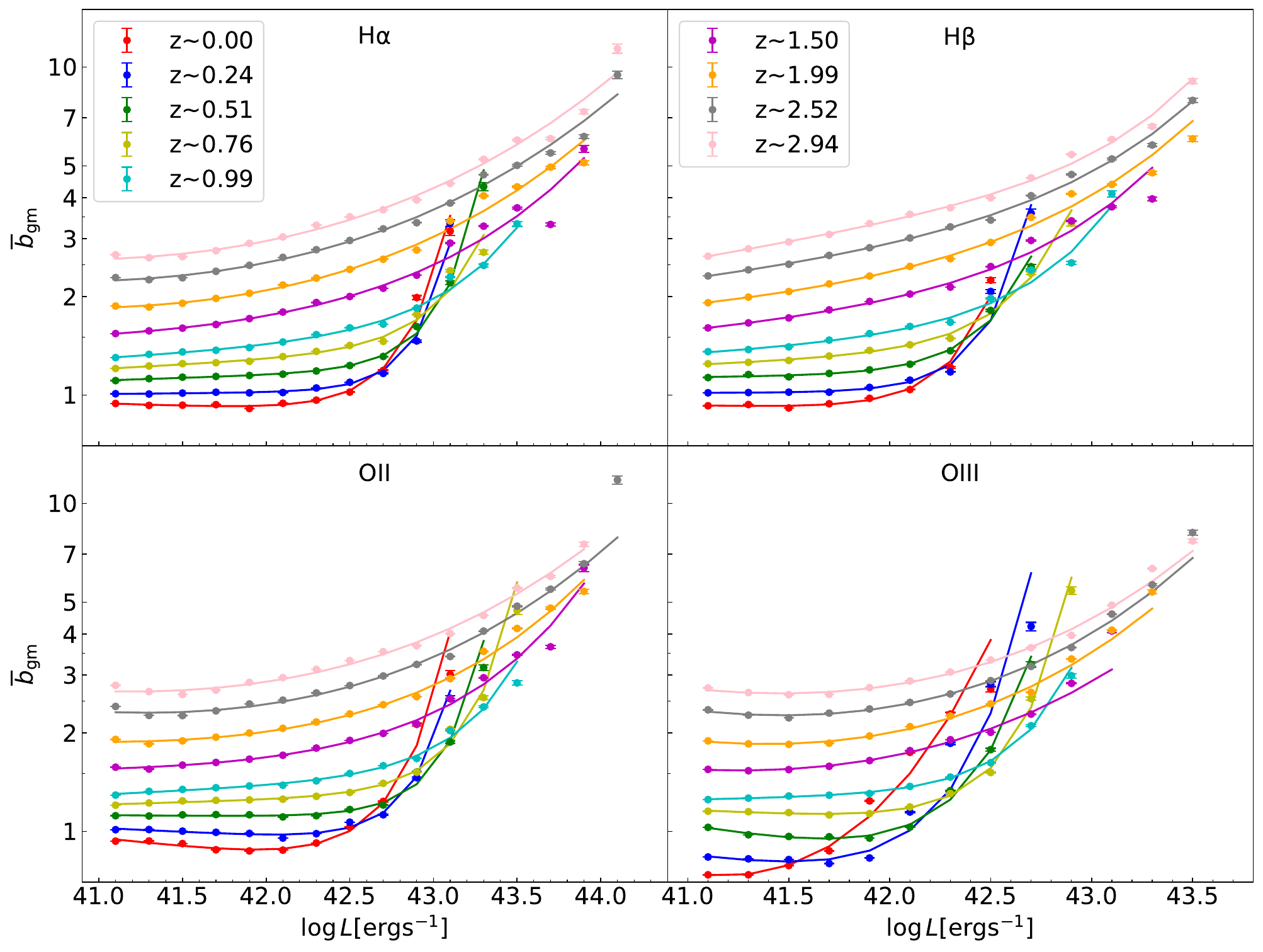}
\caption{Mean galaxy-matter bias ($\overline{b}_{\rm gm}$) as a function of luminosity from redshift 0 to 3 for H$\alpha$, H$\beta$, [OII], and [OIII] selected galaxies. Different panels represent different emission lines. Solid lines represent the fitting results using Equation~\ref{eq:biasfitting}.}
\label{fig:biasline}
\end{figure*}

\begin{figure*}
\centering
\includegraphics[width=2\columnwidth]{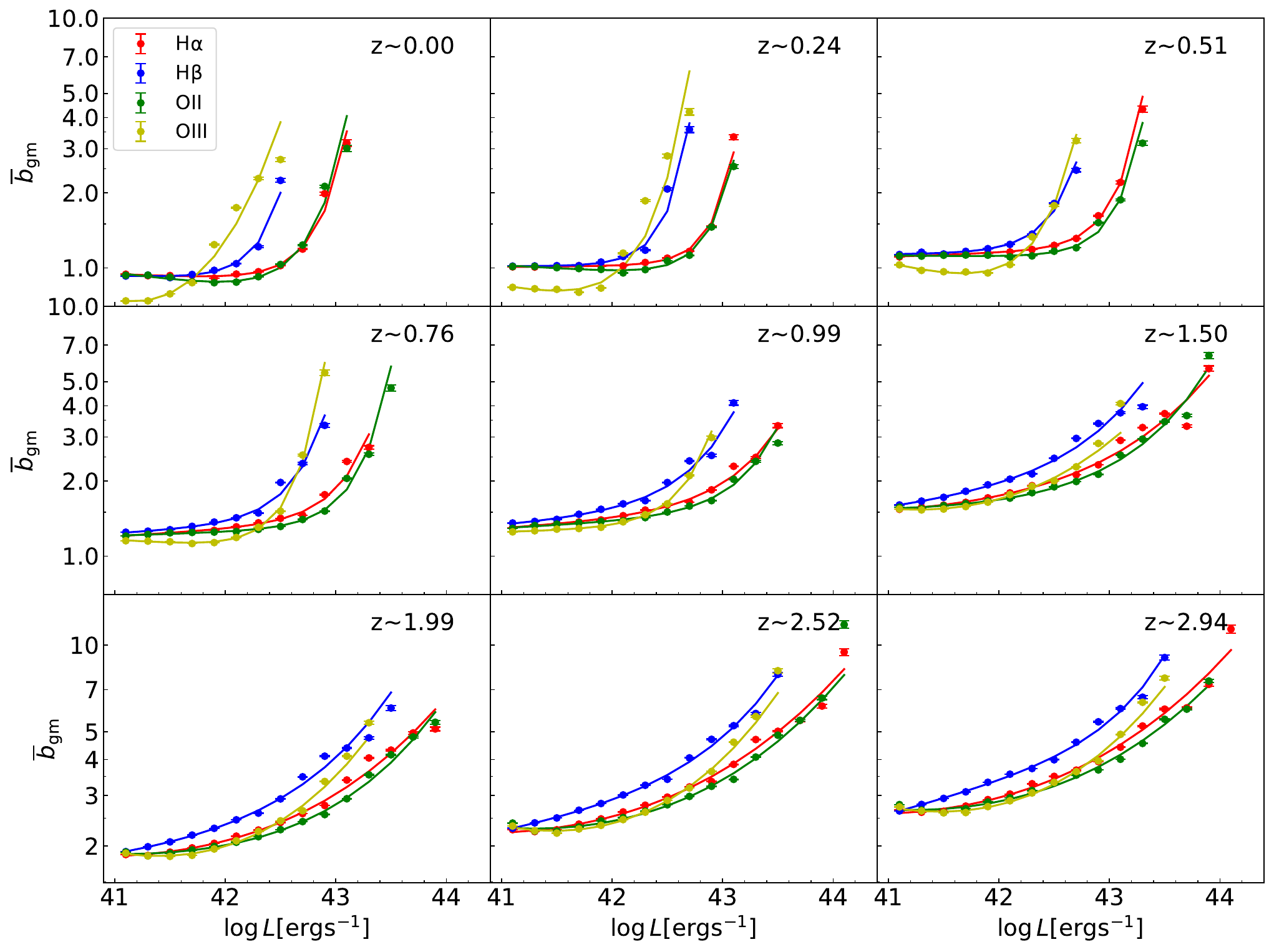}
\caption{Mean galaxy-matter bias ($\overline{b}_{\rm gm}$) as a function of luminosity for H$\alpha$, H$\beta$, [OII], and [OIII] from redshift 0 to 3. Different panels represent different redshifts. Solid lines represent the fitting results using Equation~\ref{eq:biasfitting}.}
\label{fig:bias}
\end{figure*}

\subsection{Bias of different emission line tracers}
\label{sec:emibias}
Emission line galaxies are one of the main targets for the next generation of large-scale surveys. It is essential to understand their bias relative to the matter distribution. Galaxy bias is often used in various cosmological probes, including baryonic acoustic oscillation, redshift distortion, etc. The bias is defined using :
\begin{equation}
    b_{gm}^2 = \frac{\xi_{gg}}{\xi_{mm}},
\end{equation}
where $\xi_{gg}$ is the 3D correlation function of the galaxy sample, and $\xi_{mm}$ is the correlation function of the total matter. A value of $b = 1$ signifies that the particular galaxy sample traces the distribution of the total matter, while when b is greater or less than 1, it means that the particular galaxy sample is more or less strongly clustered than the total mass. 

Here we calculate the bias of emission line galaxies as a function of luminosity and redshift. We split the simulated ELGs into bins of 0.2 dex within the luminosity range from $10^{41}$ to $10^{44} \ergs$. We focus on four of the most luminous lines, H$\alpha$, H$\beta$, [OII], and [OIII]. Bins containing fewer than 1000 galaxies are excluded. We compute the average bias over the range 20\mpch{} to 60\mpch{} where the bias is relatively constant and denote it as $\overline{b}_{\rm gm}$. Fig.~\ref{fig:biasline} presents $\overline{b}_{\rm gm}$ as a function of luminosity at different redshift 0 to 3 for H$\alpha$, H$\beta$, [OII], and [OIII] selected galaxies. The errors are Poisson errors. 

Fig.~\ref{fig:biasline} shows that for H$\alpha$ selected galaxies at z$<$1 the bias exhibits a flat slope with luminosity for less luminous objects ($<10^{42.5}\ergs$) and undergoes a sharp increase at high luminosity. The luminosity transition increases as redshifts increase and disappears at z$>$1, where the bias shows a more gradual increase with luminosity across the entire range of luminosity. At low luminosities, the bias monotonically increases with redshifts. However, at high luminosities, the bias decreases with redshifts up to z $\sim$ 1 and then increases with redshift above z$\sim$1. This is due to the sharp increase in luminosity at low redshifts which disappears at high redshifts.

Similar features are observed in galaxies selected based on other emission lines, although the exact redshift and luminosity dependence differ among different line selections. For instance, the transition luminosity is considerably smaller in galaxies selected based on [OIII] emission line compared to H$\alpha$ selected galaxies.

The variation in line selections can be understood as different line emissions corresponding to different physical conditions. Lines with the same luminosity may be hosted by galaxies with different properties. For example, for H$\alpha \sim 10^{43} \ergs$, the typical galaxy stellar mass is $10^{11.43}\Msun$, SFR is $51.26\Msun{\rm yr^{-1}}$, and halo mass is $10^{13.49}\Msun$. In contrast, for [OIII] with the same luminosity, the typical galaxy stellar mass is $10^{11.02}\Msun$, SFR is $65.81\Msun{\rm yr^{-1}}$, and halo mass is $10^{13.06}\Msun$.

Detailed comparison among different emission lines is presented in Fig.~\ref{fig:bias}. At redshift 0, it is evident that for H$\alpha$ and [OII] selected galaxies, there is almost no dependence on emission line luminosity up to $10^{42.5} \ergs$. The typical halo mass is $10^{12.6}\Msun$ for those with luminosity around the transition. However, at higher luminosities, the bias increases rapidly, reaching $\sim$ 4 at $10^{43} \ergs$, corresponding to a typical halo mass of $10^{13.5}\Msun$. 

At z=0, [OII] selected galaxies always have a similar bias compared to H$\alpha$ selected galaxies. H$\beta$ selected galaxies exhibit a comparable bias to H$\alpha$ selected and [OII] selected galaxies with luminosities below $10^{42} \ergs$. The corresponding halo mass at the transition point is $10^{12.55}\Msun$, which is also similar to the halo mass of the H$\alpha$ selected and [OII] selected galaxies. Beyond this luminosity threshold, the H$\beta$ selected galaxies experience a rapid increase in bias and surpass the bias of the H$\alpha$ selected and [OII] selected galaxies, reaching a bias of approximately 2 at $10^{42.5} \ergs$. The transition luminosity happens at even smaller luminosity for [OIII] selected galaxies.  Surprisingly, we find that for galaxies with the least luminous [OIII] emissions, the bias is well below 1, also much lower than H$\alpha$, H$\beta$ and [OII] selected galaxies of the same luminosity. The corresponding typical stellar mass is $10^{10.24}\Msun$, SFR is $3.24\Msun{\rm yr^{-1}}$, and halo mass is $10^{11.88}\Msun$. 
Such relations are also observed at other redshifts.

The relative bias of [OIII] selected galaxies compared to other line-selected galaxies varies with luminosities and redshifts. At high luminosities, H$\alpha$ and [OII] selected galaxies have a lower bias than [OIII] selected galaxies at all redshift of interest. Conversely, at low luminosities, the [OIII] selected galaxies have a lower bias below z = 1. The transition luminosity increases with increasing redshifts. At higher redshifts, [OII] selected galaxies have a similar bias compared to [OIII] selected galaxies.

For less luminous ELGs, especially at low redshifts, their bias dependence on luminosity is weak, posing challenges in modelling their spatial distribution using halo occupation models and abundance matching techniques. To quantify the relationship between $\overline{b}_{\rm gm}$ and luminosity at different redshift, we employ a combined power-law and exponential function to fit $\overline{b}_{\rm gm}(L)$:
\begin{equation}
\label{eq:biasfitting}
    \overline{b}_{\rm gm}(L) = \Phi_{\rm L} \cdot (\frac{L}{L_*}) ^ {\alpha} \cdot e ^ \frac{L}{L_*\cdot 10^{\beta}},
\end{equation}
where $\Phi_{\rm L}$, $L_*$, $\alpha$ and $\beta$ are free parameters. We set a minimum value of log($L_*$) = 40 since we do not use data from lower luminosity bins. 

The fitting results are presented in Fig.~\ref{fig:biasline} and Fig.~\ref{fig:bias} by solid lines. Our fitting formula consistently reproduces the observed bias correlation at all luminosities across all redshifts. The only exception appears at the highest luminosity end for some redshifts, where the sample size is relatively too small. The fitting parameters as a function of redshift are listed in Table~\ref{tab:fitting}. 

\begin{table*}
\centering
\begin{center}
\caption{The best-fitting parameters for $\overline{b}_{\rm gm}$ in Eq.~\ref{eq:biasfitting}.}
\label{tab:fitting}
\begin{tabular}{lcccccr}
\hline
lines & redshift & $\Phi_{\rm L}$ & $L_*$ & $\alpha$ & $\beta$ \\
\hline
H$\alpha$ & \\
& 0.00  & 0.8737 $\pm$ 0.0023  & 43.0112 $\pm$ 0.0035  & -0.0169 $\pm$ 0.0006  & 0.6110 $\pm$ 0.0077  \\
& 0.24  & 1.0291 $\pm$ 0.0016  & 43.0926 $\pm$ 0.0041  & 0.0043 $\pm$ 0.0004  & 0.4704 $\pm$ 0.0067  \\
& 0.51  & 1.2016 $\pm$ 0.0018  & 43.2209 $\pm$ 0.0030  & 0.0159 $\pm$ 0.0004  & 0.5506 $\pm$ 0.0054  \\
& 0.76  & 1.3875 $\pm$ 0.0028  & 43.3713 $\pm$ 0.0047  & 0.0266 $\pm$ 0.0004  & 0.7369 $\pm$ 0.0089  \\
& 0.99  & 1.5103 $\pm$ 0.0080  & 43.6363 $\pm$ 0.0070  & 0.0264 $\pm$ 0.0009  & 1.2152 $\pm$ 0.0219  \\
& 1.50  & 1.2460 $\pm$ 0.0318  & 43.4565 $\pm$ 0.0291  & -0.0142 $\pm$ 0.0031  & 2.7072 $\pm$ 0.0685  \\
& 1.99  & 0.7382 $\pm$ 0.0248  & 41.8121 $\pm$ 0.1817  & -0.1266 $\pm$ 0.0096  & 4.8389 $\pm$ 0.1920  \\
& 2.52  & 0.8573 $\pm$ 0.0509  & 40.0000 $\pm$ 0.4838  & -0.1993 $\pm$ 0.0193  & 6.6379 $\pm$ 0.4039  \\
& 2.94  & 0.9992 $\pm$ 0.0678  & 40.0000 $\pm$ 0.5474  & -0.2000 $\pm$ 0.0219  & 6.6310 $\pm$ 0.4566  \\
\hline
H$\beta$ & \\
& 0.00  & 0.8889 $\pm$ 0.0049  & 42.5498 $\pm$ 0.0050  & -0.0129 $\pm$ 0.0017  & 0.5425 $\pm$ 0.0131  \\
& 0.24  & 1.0183 $\pm$ 0.0030  & 42.6426 $\pm$ 0.0033  & 0.0007 $\pm$ 0.0009  & 0.4824 $\pm$ 0.0069  \\
& 0.51  & 1.1557 $\pm$ 0.0042  & 42.7490 $\pm$ 0.0041  & 0.0049 $\pm$ 0.0010  & 0.5996 $\pm$ 0.0096  \\
& 0.76  & 1.3819 $\pm$ 0.0055  & 42.9082 $\pm$ 0.0040  & 0.0262 $\pm$ 0.0010  & 0.7252 $\pm$ 0.0106  \\
& 0.99  & 1.5713 $\pm$ 0.0130  & 43.1604 $\pm$ 0.0064  & 0.0345 $\pm$ 0.0016  & 1.0849 $\pm$ 0.0246  \\
& 1.50  & 1.9400 $\pm$ 0.0330  & 43.3461 $\pm$ 0.0087  & 0.0454 $\pm$ 0.0025  & 1.6593 $\pm$ 0.0420  \\
& 1.99  & 1.8870 $\pm$ 0.0632  & 43.2523 $\pm$ 0.0303  & 0.0209 $\pm$ 0.0042  & 2.3328 $\pm$ 0.0744  \\
& 2.52  & 3.0811 $\pm$ 0.0749  & 43.5481 $\pm$ 0.0138  & 0.0607 $\pm$ 0.0030  & 1.9206 $\pm$ 0.0641  \\
& 2.94  & 4.8227 $\pm$ 0.0455  & 43.7044 $\pm$ 0.0108  & 0.1014 $\pm$ 0.0015  & 1.2937 $\pm$ 0.0396  \\
\hline
OII & \\
& 0.00  & 0.7444 $\pm$ 0.0020  & 42.9343 $\pm$ 0.0022  & -0.0561 $\pm$ 0.0007  & 0.7061 $\pm$ 0.0061  \\
& 0.24  & 0.9096 $\pm$ 0.0015  & 43.0808 $\pm$ 0.0028  & -0.0251 $\pm$ 0.0004  & 0.5507 $\pm$ 0.0056  \\
& 0.51  & 1.0983 $\pm$ 0.0014  & 43.2477 $\pm$ 0.0025  & -0.0041 $\pm$ 0.0003  & 0.5527 $\pm$ 0.0047  \\
& 0.76  & 1.3020 $\pm$ 0.0018  & 43.3889 $\pm$ 0.0025  & 0.0137 $\pm$ 0.0003  & 0.6514 $\pm$ 0.0049  \\
& 0.99  & 1.4827 $\pm$ 0.0034  & 43.5863 $\pm$ 0.0045  & 0.0230 $\pm$ 0.0004  & 0.9035 $\pm$ 0.0101  \\
& 1.50  & 1.4274 $\pm$ 0.0142  & 43.6284 $\pm$ 0.0064  & -0.0068 $\pm$ 0.0014  & 1.9014 $\pm$ 0.0269  \\
& 1.99  & 1.1036 $\pm$ 0.0253  & 43.1303 $\pm$ 0.0382  & -0.0672 $\pm$ 0.0034  & 3.0528 $\pm$ 0.0636  \\
& 2.52  & 0.9873 $\pm$ 0.0828  & 40.0000 $\pm$ 0.3568  & -0.2549 $\pm$ 0.0164  & 6.2921 $\pm$ 0.2918  \\
& 2.94  & 1.1106 $\pm$ 0.1041  & 40.0000 $\pm$ 0.4619  & -0.2372 $\pm$ 0.0198  & 6.4710 $\pm$ 0.3805  \\
\hline
OIII & \\
& 0.00  & 0.5453 $\pm$ 0.1945  & 40.0000 $\pm$ 0.3073  & -0.7762 $\pm$ 0.0771  & 3.0961 $\pm$ 0.2177  \\
& 0.24  & 0.6130 $\pm$ 0.0050  & 42.3960 $\pm$ 0.0044  & -0.0969 $\pm$ 0.0026  & 0.8110 $\pm$ 0.0099  \\
& 0.51  & 0.6976 $\pm$ 0.0042  & 42.5272 $\pm$ 0.0026  & -0.1119 $\pm$ 0.0017  & 0.8129 $\pm$ 0.0096  \\
& 0.76  & 1.0328 $\pm$ 0.0026  & 42.7483 $\pm$ 0.0020  & -0.0294 $\pm$ 0.0007  & 0.6180 $\pm$ 0.0054  \\
& 0.99  & 1.3090 $\pm$ 0.0038  & 42.9383 $\pm$ 0.0036  & 0.0106 $\pm$ 0.0007  & 0.6964 $\pm$ 0.0088  \\
& 1.50  & 0.6956 $\pm$ 0.1041  & 40.0000 $\pm$ 0.4643  & -0.3188 $\pm$ 0.0292  & 5.3731 $\pm$ 0.3808  \\
& 1.99  & 0.9224 $\pm$ 0.1064  & 40.0000 $\pm$ 0.2736  & -0.3689 $\pm$ 0.0203  & 5.0910 $\pm$ 0.2173  \\
& 2.52  & 1.1697 $\pm$ 0.1324  & 40.0000 $\pm$ 0.2544  & -0.3761 $\pm$ 0.0188  & 5.1419 $\pm$ 0.2002  \\
& 2.94  & 1.3221 $\pm$ 0.1650  & 40.0000 $\pm$ 0.3063  & -0.3542 $\pm$ 0.0209  & 5.3237 $\pm$ 0.2422  \\
\hline
\end{tabular}
\end{center}
\end{table*}

In summary, in combination with the SAM and emission line models, we successfully reproduce most of the observed properties of emission line galaxies properties, including their luminosity functions, correlation functions, and their evolution with redshifts. At high redshifts, their bias increases with both redshift and luminosity, while at low redshift, the bias is a decreasing function of redshift, attributed to the stronger dependence on luminosity towards lower redshifts. Across all redshifts, the luminosity dependence of galaxy bias is weak below $10^{42} \ergs$

\section{Conclusions}
\label{sec:conc}

Our investigation focuses on generating a simulated galaxy catalogue for next-generation surveys, especially to include emission line galaxies in a self-consistent way. We solve the time convergence issue in the widely used semi-analytic model \lgalaxies{} and employ it in the JiuTian-1G simulation. Furthermore, we compute the luminosity of various emission lines, enabling predictions about emission line galaxies. We further study the clustering and bias of different emission lines and provide a fitting formula for bias as a function of luminosity and redshift. Our catalogue successfully reproduces various observational properties. The main conclusions of our study can be summarized as follows:

\begin{enumerate}
 	\item We observe a significant convergence problem in the \lgalaxies{} model presented by \citet{2015MNRAS.451.2663H} when applied to dark matter merger trees with varying time intervals. This issue stems from the disruption model being exclusively implemented at the end of each snapshot gap. Therefore, merger trees with fewer snapshots tend to experience reduced disruption and an increased number of mergers. As mergers predominantly contribute to the growth of SMBH, a higher frequency of mergers results in more massive SMBHs, leading to more efficient AGN feedback and, consequently, less massive galaxies. By modifying the disruption model in \lgalaxies{}, We successfully achieve excellent convergence in simulations with merger trees of varying time intervals.
        \\[-.3cm]
      \item Our adapted model is applied to the JiuTian-1G simulation, and the corresponding parameters were readjusted.  
      Our catalogue successfully reproduces numerous statistical observational properties and accurately captures the clustering patterns of diverse galaxy samples. In particular, it has been able to replicate the SMBH mass function, which was underestimated by H15 by an order of magnitude.
      
      \item We demonstrate that in combination with the high resolution large boxsize JT1G simulation, \lgalaxies{} has successfully generated a galaxy catalogue that fulfils most of the requirements of next-generation large-scale surveys.

      \item We compute the luminosity of 13 commonly used NUV and optical emission lines. Our model effectively reproduces the observed luminosity function of H$\alpha$, H$\beta$, [OII], and [OIII]. Additionally, the projected correlation of [OII] ELGs shows good agreement with observations.  
      \item We further explore the bias of emission line galaxies as a function of luminosity and redshift. The dependence varies with luminosity ranges and redshifts. We observe that at low redshift, the bias of galaxies with low luminosity shows insensitivity to luminosity, while it increases rapidly at the high luminosity end. At high redshift, bias gradually increases with luminosity. Above z=1, galaxy bias increases monotonically with redshift, while below z=1, such a monotonic increase only holds for low luminosity galaxies. At high luminosities, galaxy bias decreases with redshift up to z=1 and then increases with redshift.
     \item We offer fitting formulas that capture the dependence of bias on both luminosity and redshifts.
\end{enumerate}

In conclusion, our adapted model successfully replicates various galaxy observational properties, and the predictions from our emission line model align well with observations. The bias has a complex dependence on luminosity and redshift, which varies with luminosity range and redshift range.

Due to the limitation of storage, we only present photometric magnitude from several surveys and the luminosity of 13 emission lines. Additional photometric magnitude, emission lines, and full SED are available upon request.

\section*{Acknowledgements}

We thank Professor Simon White for his suggestions. This work is supported by the National SKA Program of China (Nos. 2022SKA0110201 and 2022SKA0110200), and CAS Project for Young Scientists in Basic Research grant No. YSBR-062, the National Natural Science Foundation of China (NSFC) (grant Nos. 12033008, 12273053, 11988101, 11622325), the K.C.Wong Education Foundation, the science research grants from China Manned Space Project with No.CMS-CSST-2021-A03 and No.CMS-CSST-2021-B03, and the Strategic Priority Research Program of Chinese Academy of Sciences, Grant No.XDB0500203. QG thanks European Union’s HORIZON-MSCA-2021-SE-01 Research and Innovation programme under the Marie Sklodowska-Curie grant agreement number 101086388. JH acknowledges support from the Yangyang development fund.

\section*{Data Availability}

The data produced in this paper are available upon reasonable request to the corresponding author.



\bibliographystyle{mnras}
\bibliography{ref} 




\appendix

\section{Observational Constraints used in MCMC}

In the MCMC procedure, we use the stellar mass function at z = 0, 1, 2, and 3, the passive fraction at z = 0.4, and the black hole mass function at z = 0 as constraints to find the best-fitting parameters.
The observational data we use in this study are listed in Table~\ref{tab:obs}. The constraints of the SMF at z = 0,1,2, and 3 are combined from the listed data-set \citep[see Appendix A in][for details]{2015MNRAS.451.2663H}, while the passive fraction and BHMF only rely on data from a single observation.

\begin{table}
\centering
  \caption{Observed data set we use in MCMC procedures. The constraints of SMF at z = 0, 1, 2, and 3 are combined from the listed data set, while the passive fraction and BHMF only use data from one observation.}
\label{tab:obs}
\begin{tabular}{lcr}
\hline
\hline
 constraints & Publication \\
\hline
\hline
SMF, $z=0.0$ & \citet{2008MNRAS.388..945B},\\ & \citet{2009MNRAS.398.2177L},\\ & \citet{2012MNRAS.421..621B} \\
\hline
SMF, $z=1.0$ & \citet{2010ApJ...709..644I},\\ & \citet{2013AA...556A..55I},\\ & \citet{2013ApJ...777...18M},\\ & \citet{2014ApJ...783...85T} \\
\hline
SMF, $z=2.0$ & \citet{2011MNRAS.417..900D},\\ & \citet{2013AA...556A..55I},\\ & \citet{2013ApJ...777...18M},\\ & \citet{2014ApJ...783...85T}\\
\hline
SMF, $z=3.0$ & \citet{2009ApJ...701.1765M},\\ & \citet{2010ApJ...725.1277M},\\ & \citet{2011MNRAS.417..900D},\\ & \citet{2013AA...556A..55I},\\ & \citet{2013ApJ...777...18M} \\
\hline
\hline
$f_{\rm passive}$, $z=0.4$ & \citet{2010ApJ...709..644I}\\
\hline
\hline
BHMF, $z=0.0$ & \citet{2009ApJ...690...20S} \\
\hline
\hline
\end{tabular}
\end{table}

\section{Convergence of our new model}
\label{app:conv}

\begin{figure*}
\centering
\includegraphics[width=2\columnwidth]{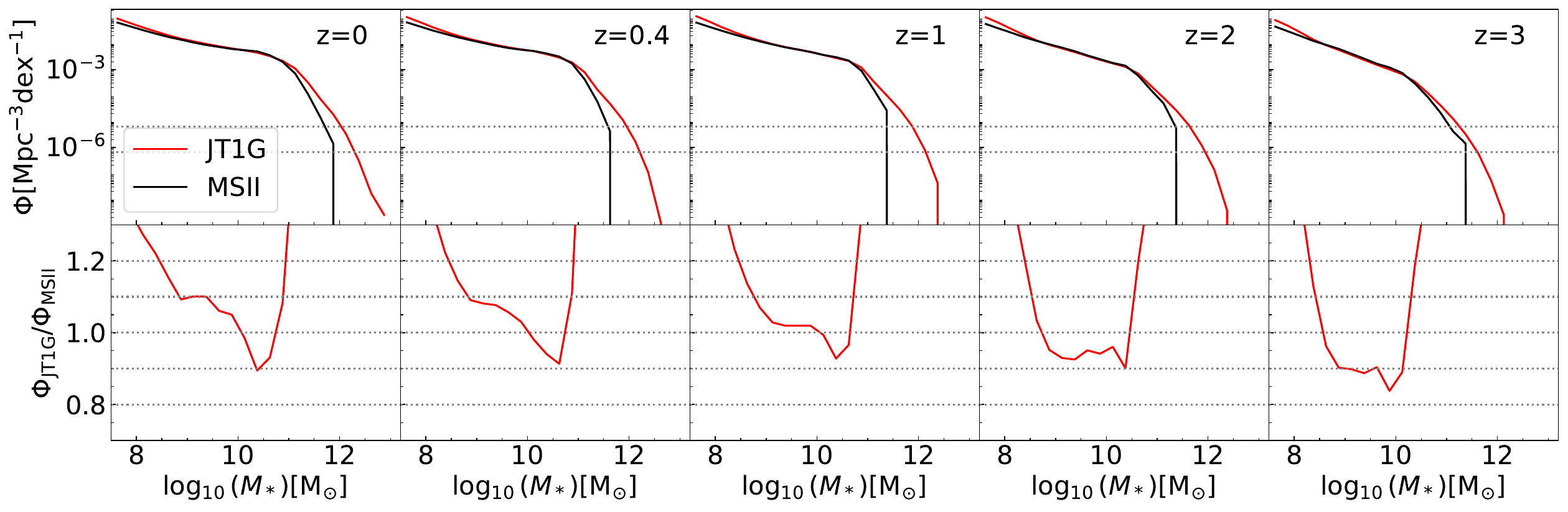}
\caption{Performance of our model in predicting the stellar mass function (SMF) from redshift 0 to 3 on both the JT1G and MSII simulations. The red curves represent the JT1G simulation, while the black curves represent the MSII simulation. The upper row depicts the SMF, and the lower row illustrates the ratio of JT1G to MSII. The two horizontal dashed lines in the upper panels represent one galaxy and ten galaxies in MSII, respectively.}
\label{fig:conv_MSII_SMF}
\end{figure*}

\begin{figure*}
\centering
\includegraphics[width=2\columnwidth]{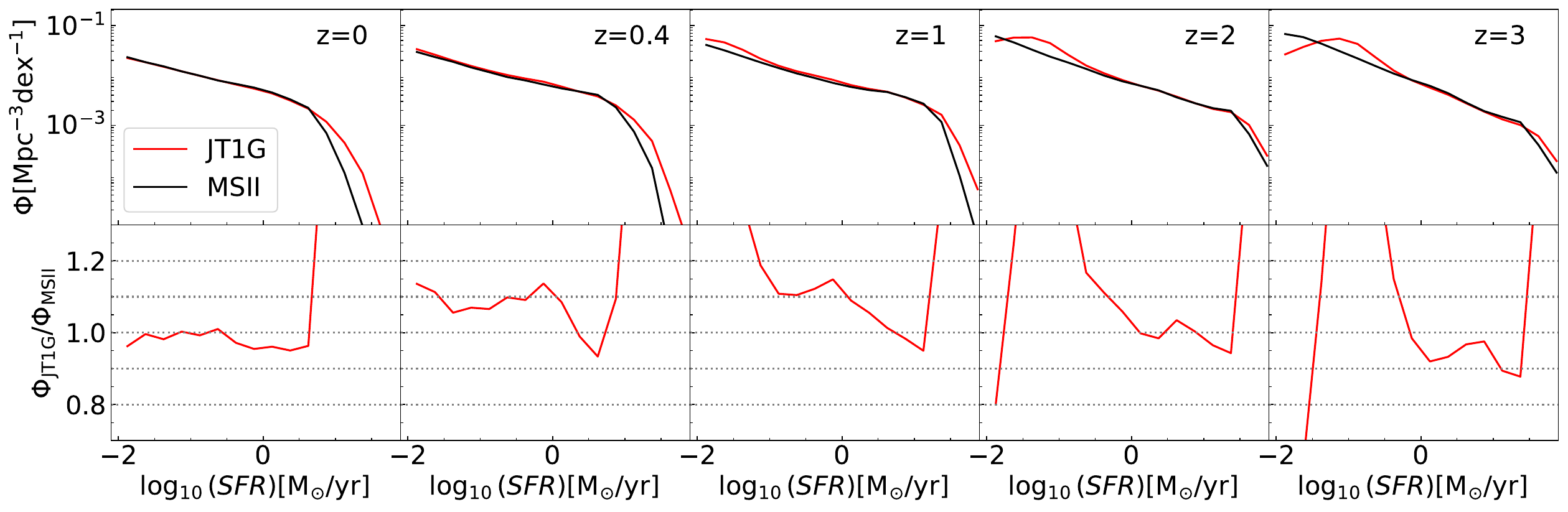}
\caption{Same as Fig.~\ref{fig:conv_MSII_SMF}, but for SFRF.}
\label{fig:conv_MSII_SFRF}
\end{figure*}

In order to assess the convergence of our model across simulations with varying dark matter particle masses, we conducted our analyses on the high-precision re-scaled Millennium-II (MSII) simulation \citep{2009MNRAS.398.1150B}. The dark matter particle mass in the MSII simulation is set to $7.69\times10^{6}\Msunh$, with precision nearly 50 times greater than that of the JT1G simulation. Furthermore, the MSII simulation has been scaled to the cosmological parameters of Planck14, which differ slightly from those used in our Planck18-based model. Therefore, comparing the performance of our model across the MSII and JT1G simulations allows for a more nuanced understanding of its convergence concerning varying resolutions and cosmological parameters. Fig.~\ref{fig:conv_MSII_SMF} illustrates the performance of our model in predicting the SMF from redshift 0 to 3 on both the JT1G and MSII simulations. The red curves represent the JT1G simulation, while the black curves represent the MSII simulation. The upper row depicts the SMF, and the lower row illustrates the ratio of JT1G to MSII. Overall, our model exhibits a robust agreement between JT1G and MSII within the range of $10^{9}\Msun$ to $10^{10.5}\Msun$, with differences consistently below 10\%. At the lower mass end, there is a noticeable overestimation due to the precision limitations of the JT1G simulation. Conversely, at the higher mass end, the MSII simulation, constrained by its box volume, lacks massive galaxies. 

Fig.~\ref{fig:conv_MSII_SFRF} is the same as Fig.~\ref{fig:conv_MSII_SMF}, but presenting the Star Formation Rate Function (SFRF) across redshifts 0 to 3. At lower redshifts, the agreement between JT1G and MSII results is excellent, with a deviation of less than 5\%. However, at higher redshifts, JT1G deviates significantly from MSII at lower SFR values, indicating a limitation in resolution. Notably, at SFR values greater than 1$\Msun{\rm yr^{-1}}$, JT1G and MSII results remain in close agreement.

In summary, our model demonstrates outstanding temporal resolution convergence and satisfactory convergence across simulations with different particle masses. The JT1G simulation provides comprehensive and resolution-independent coverage for galaxies with stellar masses above $10^{9}\Msun$ and SFR exceeding 1$\Msun{\rm yr^{-1}}$, up to redshift 3.

\section{SEDs of two typical galaxies}
\label{app:sed}

\begin{figure}
\includegraphics[width=\columnwidth]{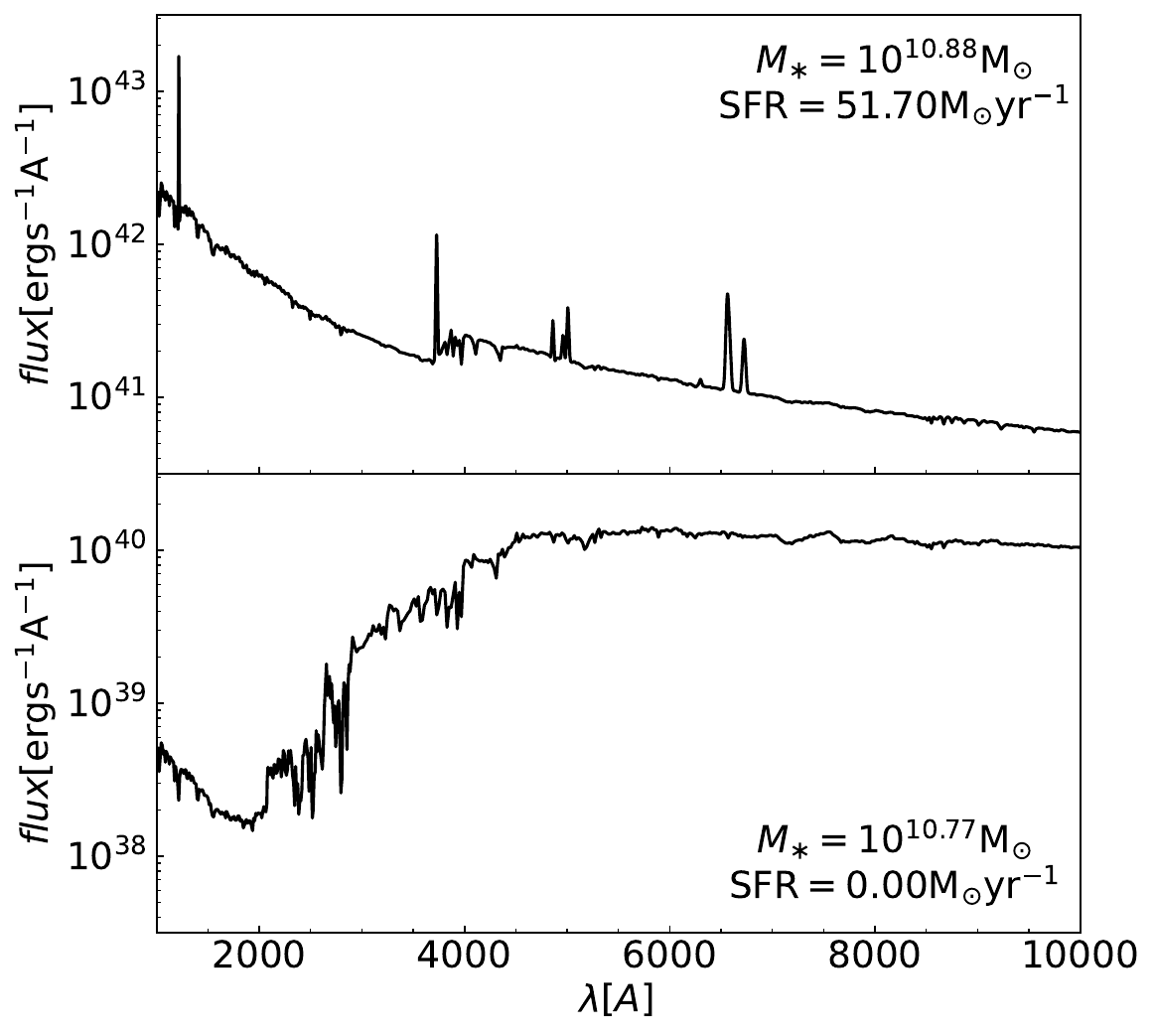}
\caption{Upper panel: Full SED of a typical star-forming galaxy in our catalogue with emission lines. The stellar mass is $10^{10.88}\Msun$ and the star formation rate is 51.70${\rm \Msun yr^{-1}}$. Bottom panel: Full SED of a typical quenched galaxy in our catalogue. The stellar mass is $10^{10.77}\Msun$ and the star formation rate is 0${\rm \Msun yr^{-1}}$.}
\label{fig:sed}
\end{figure}

In general, we can calculate the galaxy SED with emission lines for each galaxy in the catalogue. Here, the SEDs of a typical star-forming galaxy and a passive galaxy are presented in Fig.~\ref{fig:sed} using the method described in Secion~\ref{sec:elm}. We include the 13 emission lines in the SED for star-forming galaxies. It illustrates the declining feature with increasing wavelength for the star-forming galaxy, the D4000 break and UV-upturn for the passive galaxy.


\bsp	
\label{lastpage}
\end{document}